\documentclass[acmsmall,authorversion]{acmart}

\usepackage{multirow}
\usepackage{graphicx}
\usepackage{grffile}
\usepackage{listings}
\usepackage[frozencache,cachedir=.]{minted}

\setminted{
  autogobble,
  breaklines,
  fontsize  = \scriptsize,
  frame     = single,
  linenos,
  numbersep = 0.5ex,
}

\definecolor{peekcolor}{rgb}{0.760, 0.890, 1}
\definecolor{readcolor}{gray}{0.888}
\newcommand{\colorpeek}[1]{{\setlength{\fboxsep}{1pt}\colorbox{peekcolor}{#1}}}
\newcommand{\colorread}[1]{{\setlength{\fboxsep}{1pt}\colorbox{readcolor}{#1}}}

\AtBeginDocument{%
  }

\acmJournal{TRETS}
\acmVolume{1}
\acmNumber{1}
\acmArticle{1}
\acmMonth{8}

\newcommand{\toolName}{TAPA}

\begin{document}

\title{\toolName: A Scalable Task-Parallel Dataflow Programming Framework for Modern FPGAs with Co-Optimization of HLS and Physical Design}

\author{Licheng Guo}
\authornote{indicates equal contribution.}
\email{lcguo@cs.ucla.edu}
\author{Yuze Chi}
\authornotemark[1]
\email{chiyuze@cs.ucla.edu}
\author{Jason Lau}
\authornotemark[1]
\email{lau@cs.ucla.edu}
\author{Linghao Song}
\email{linghaosong@cs.ucla.edu}
\affiliation{%
  \institution{University of California Los Angeles}
  \city{Los Angeles}
  \state{California}
  \country{USA}
}

\author{Xingyu Tian}
\email{xingyu\_tian@sfu.ca}
\author{Moazin Khatti}
\email{moazin\_khatti@sfu.ca}
\affiliation{%
  \institution{Simon Fraser University}
  \city{Burnaby}
  \city{British Columbia}
  \country{Canada}
}

\author{Weikang Qiao}
\email{wkqiao2015@ucla.edu}
\author{Jie Wang}
\email{jaywang@cs.ucla.edu}

\affiliation{%
  \institution{University of California Los Angeles}
  \city{Los Angeles}
  \state{California}
  \country{USA}
}

\author{Ecenur Ustun}
\email{eu49@cornell.edu}
\affiliation{%
  \institution{Cornell University}
  \city{Ithaca}
  \state{New York}
  \country{USA}
}


\author{Zhenman Fang}
\email{zhenman@sfu.ca }
\affiliation{%
  \institution{Simon Fraser University}
  \city{Burnaby}
  \city{British Columbia}
  \country{Canada}
}

\author{Zhiru Zhang}
\email{zhiruz@cornell.edu}
\affiliation{%
  \institution{Cornell University}
  \city{Ithaca}
  \state{New York}
  \country{USA}
}

\author{Jason Cong}
\email{cong@cs.ucla.edu}

\affiliation{%
  \institution{University of California Los Angeles}
  \city{Los Angeles}
  \state{California}
  \country{USA}
}

\renewcommand{\shortauthors}{L. Guo et al.}

\begin{abstract}

In this paper, we propose TAPA, an end-to-end framework that compiles a C++ task-parallel dataflow program into a high-frequency FPGA accelerator. Compared to existing solutions, TAPA has two major advantages. First, TAPA provides a set of convenient APIs that allow users to easily express flexible and complex inter-task communication structures. Second, TAPA adopts a coarse-grained floorplanning step during HLS compilation for accurate pipelining of potential critical paths. In addition, TAPA implements several optimization techniques specifically tailored for modern HBM-based FPGAs. In our experiments with a total of 43 designs, we improve the average frequency from 147 MHz to 297 MHz (a 102\% improvement) with no loss of throughput and a negligible change in resource utilization. Notably, in 16 experiments we make the originally unroutable designs achieve 274 MHz on average. The framework is available at \textit{https://github.com/UCLA-VAST/tapa} and the core floorplan module is available at \textit{https://github.com/UCLA-VAST/AutoBridge}.

\end{abstract}



\maketitle


\section{Introduction}

High-level synthesis (HLS) with the task-parallel programming model is an important tool to help programmers scale up the performance of their accelerators on modern FPGAs with ever-increasing resource capacities. Task-level parallelism is a form of parallelization of computer programs across multiple processors. In contrast to data parallelism where the workload is partitioned on data and each processor executes the same program (e.g., OpenMP~\cite{cse98-openmp}), different processors in a task-parallel program often behave differently, while data are passed between processors. Examples of task-parallel programs include image processing pipelines~\cite{taco17-halide-hls, iccad18-soda, fpga20-heterohalide}, graph processing~\cite{fpga16-fpgp, fpga17-foregraph, tpds19-hitgraph, fccm19-worklist-graph} and network switching~\cite{fpga20-switch}. Researches show that even for data-parallel applications like neural networks~\cite{wang2021autosa, sohrabizadeh2020end, sohrabizadeh2022streamgcn} and stencil computation~\cite{iccad18-soda}, task-parallel implementations show better scalability and higher frequency than their data-parallel counterparts due to the localized communication pattern~\cite{fccm18-latte}. 

Even though task-parallel programs are suitable for spatial architectures, existing FPGA computer-aided design (CAD) toolchains still often fail in timing closure. One major cause that leads to the unsatisfactory frequency is that HLS cannot easily predict the physical layout of the design after placement and routing, thus HLS tools typically rely on pre-characterized operation delays and a very crude interconnect delay model to insert clock boundaries (i.e., registers) into an untimed design to generate a timed RTL implementation~\cite{zheng2014fast,tan2015mapping,guo2020analysis}. Hence, as the HLS designs get larger, the behavior-level estimation becomes even less accurate and the timing quality of the synthesized RTLs usually further degrades. 

This timing issue is worsened as modern FPGA architectures become increasingly heterogeneous~\cite{url-ultrascale}. Modern FPGAs have thousands of heterogeneous digital signal processing (DSP) and random-access memory (RAM) blocks and millions of lookup table (LUT) and flip-flop (FF) instances. In order to pack more logic onto a single device, the latest FPGAs integrate multiple dies using silicon interposers, but the interconnects that go across the die boundaries would carry a non-trivial delay penalty. In addition, specialized IP blocks such as PCIe and DDR controllers are embedded in the programmable logic. These IP blocks usually have fixed locations near-dedicated I/O banks and will consume a large number of programmable resources nearby. As a result, these dedicated IPs often  detour the signals close by towards more expensive and/or longer routing paths. This complexity and heterogeneity significantly challenge the effectiveness and efficiency of modern FPGA CAD workflow.

Moreover, the recent release of High Bandwidth Memory (HBM) based FPGA boards brings even more challenges to the timing closure of HLS designs. The key feature of the HBM device is that a super wide data interface is exposed in a local region, which often causes severe local congestion. For example, the AMD/Xilinx U280 FPGA features 32 independent HBM channels at the bottom of the device, each with a 256-bit width running at 450 MHz. To fully utilize the potential of the HBM bandwidth, the physical design tools need to squeeze a substantial amount of logic into the area nearby the HBM blocks.

In this paper, we identify that the current timing-closure struggle originates from a disconnection between the HLS step and the downstream physical design step. The existing FPGA compilation stack consists of a sequence of independent design and optimization steps to gradually lower the design abstraction, but these steps lack cross-stack coordination and optimizations.
Given a C++ task-parallel program input, HLS has the flexibility to adjust the output RTL to change the pipeline level of the inter-task data links, which are usually latency-insensitive; however, the tool does not know which ones will be timing critical. On the other hand, the physical design tools could determine the critical paths but no longer have the option to add extra registers because physical design tools honor the cycle-accurate RTL input.


Therefore, we propose to re-structure the CAD stack and partially combine physical design with HLS. Specifically, we propose to couple the coarse-grained floorplanning step with behavior-level pipelining in HLS. Our coarse-grained floorplanning involves dividing the FPGA device into a grid of regions and assigning each task to one region during HLS compilation. 
We further pipeline all the inter-region connections to facilitate timing closure while we leave the intra-region optimization to the default HLS tool. As our experiment will show, floorplanning a 4-die FPGA into only 6-8 regions is already enough to properly guide HLS for accurate elimination of global critical paths, thus our floorplan-guided HLS approach is lightweight and highly scalable. 

Our methodology relieves local congestion and fixes global critical paths at the same time. First, the early floorplanning step could guide the subsequent placement steps to distribute the user logic evenly across the entire device instead of attempting to pack the logic into a single die as much as possible, which aims to alleviate local congestion as much as possible.
Second, the floorplan provides HLS a view of the global physical layout which helps HLS accurately identify and pipeline the long wires, especially those crossing the die boundaries, so the global critical paths could be appropriately pipelined. Finally, we present analysis and latency balancing algorithms to guarantee that the throughput of the resulting design is not negatively impacted. Our contributions are as follows:


\begin{itemize}

    \item To the best of our knowledge, we are the first to tackle the challenge of high-frequency HLS design on multi-die FPGAs by coupling floorplanning and pipelining to effectively insert registers on the long cross-die interconnects. We further ensure that the additional latency does not affect the throughput of the design.

    \item We present a set of optimizations specifically tailored for HBM devices, including automatic HBM port binding, floorplan solution space exploration, and a customized programming API to minimize the area overhead of HBM IO modules.
    
    \item Our framework, TAPA, interfaces with the commercial FPGA design tool flow. It improves the average frequency of 43 designs from 147 MHz to 297 MHz with a negligible area overhead. 
\end{itemize}

\begin{figure*}[!h]
	\centering
	\includegraphics[width=14.5cm]{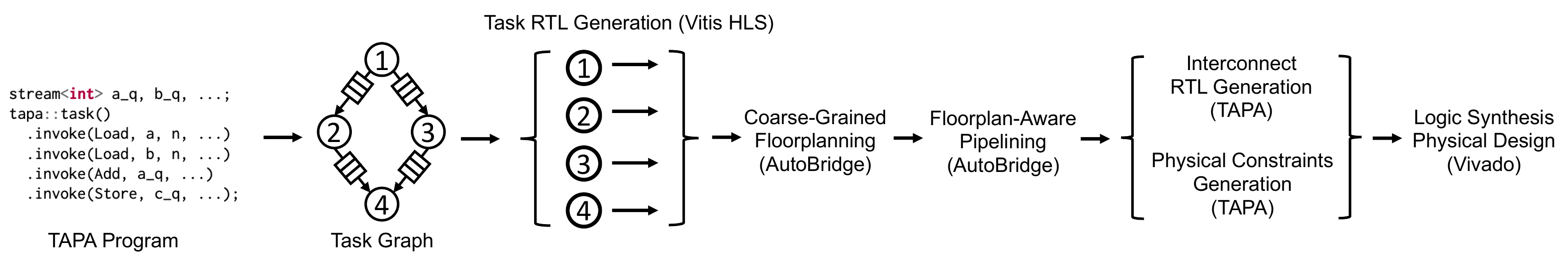}
	\caption{An overview of our TAPA framework. The input is a task-parallel dataflow program written in C/C++ with the TAPA APIs. We first invoke the TAPA compiler to extract the parallel tasks and synthesize each task using Vitis HLS to get its RTL representation and obtain an estimated area. Then the AutoBridge~\cite{fpga21-autobridge} module of TAPA floorplans the program and determines a target region for each task. Based on the floorplan, we intelligently compute the pipeline stages of the communication logic between tasks and ensure that throughput will not degrade. TAPA generates the actual RTL of the pipeline logic that composes together the tasks. A constraint file is also generated to pass the floorplan information to the downstream tools.}
    \label{fig:flow}
\end{figure*}

This paper extends the two prior publications~\cite{fpga21-autobridge, chi2020extending} of the authors on this topic. Compared to the prior papers, this paper includes additional contributions as follows:
\begin{itemize}
    \item We integrate the co-optimization methodology from~\cite{fpga21-autobridge} with the programming framework in~\cite{chi2020extending} to provide a fully automated, programmer-friendly, and robust workflow that consistently achieves higher frequency compared to existing commercial toolchains. 

    \item We extend the framework of~\cite{chi2020extending} with additional APIs for external memory access, which has significantly lowered BRAM consumption. This optimization enables the successful implementation of large-scale accelerators onto modern HBM-based FPGAs (Section~\ref{sec:async-mmap}).

    \item We extend the co-optimization methods of~\cite{fpga21-autobridge} to consider the special challenges of programming HBM-based FPGAs, including automatic HBM channel binding and multi-floorplanning generation (Section~\ref{sec:opt-hbm}).

    \item We add four extra benchmarks that use a large number of HBM channels. We demonstrate how our new optimization enables them to be successfully implemented with high frequency (Section~\ref{sec:hbm-expr}).

\end{itemize}

Figure~\ref{fig:flow} shows the overall flow of our proposed methodology.  The rest of the paper is organized as follows: Section~\ref{sec:background} introduces background information on modern FPGA architectures and shows motivating examples; Section~\ref{sec:programming-model} presents our proposed task-parallel programming model; 
 Section~\ref{sec:floorplan} details our coarse-grained floorplan scheme inside the HLS flow; Section~\ref{sec:steal} describes our floorplan-aware pipelining methods;  Section~\ref{sec:opt-hbm} elaborates our techniques tailored for HBM-based FPGAs; Section~\ref{sec:expriments} presents experimental results; Section~\ref{sec:related} provides related work, followed by conclusion and acknowledgements.

\section{Background and Motivating Examples}
\label{sec:background}


\subsection{High-Level Synthesis}

The rapid increase of complexity in FPGA design has pushed the industry and academia to raise the design abstractions with better productivity than the register transfer level (RTL). High-level synthesis (HLS) plays a central role by enabling the automatic synthesis of high-level, untimed, or partially timed specifications (e.g. C++ or OpenCL) to low-level cycle-accurate RTL specifications for efficient implementation in the field programmable gate arrays (FPGAs) or application-specific integrated circuits (ASICs). The typical flow of modern FPGA HLS tools usually consists of three core steps: (1) scheduling, (2) binding, and (3) RTL generation.

\begin{itemize}
    \item The \textit{scheduling} phase inserts clock boundaries into the original untimed specification. It takes in the control data flow graph (CDFG) generated by a compiler front-end from the high-level description (e.g., C/C++) and then maps the operations in the CDFG to the states and the control flow to state transitions specified by a finite-state machine (FSM). In each clock cycle, the controller would be in a state in the corresponding FSM.

    \item The \textit{binding} process maps high-level operations and variables to RTL-level resources, such as functional units and registers. It maps variables to registers and links wires from registers to functional units as operands of operations. The result of a functional unit is then wired to another functional unit or a register to store the computed value.
    
    \item The \textit{RTL generation} phase creates concrete RTL based on the results of the scheduling and the binding steps. The key in this step is to properly create the control logic to orchestrate the datapath, controlling each stage to execute at its scheduled cycle. 
    
\end{itemize}

\subsection{Task-Parallel Dataflow Programs}

Task-level parallelism is a form of parallelization of computer programs across
  multiple processors.
In contrast to data parallelism where the workload is partitioned on data
  and each processor executes the same program
  (e.g., OpenMP~\cite{cse98-openmp}),
  different processors in a task-parallel program often behave differently,
  while data are passed between processors.
For example, the multiple stages in an image processing pipeline~\cite{taco17-halide-hls, iccad18-soda,
  fpga20-heterohalide} can each be implemented in a different processor.  
Task-parallel programs are often described using dataflow
  models~\cite{cacm78-csp, ifip74-kpn, ieee87-sdf, thesis93-boolean-dataflow,
    acmcs77-petrinet},
  where tasks are called \textit{processes}.
Processes communicate only through unidirectional \textit{channels}.
Data exchanged between channels are called \textit{tokens}.
In this paper,
  we borrow the terms \textit{channel} and \textit{token},
  and focus on the problem of statically mapping tasks to hardware.
That is,
  instances of tasks are synthesized to different areas in an FPGA accelerator.
  
\subsection{Multi-Die FPGA Architectures}

\begin{figure}[!t]
	\centering
	\includegraphics[width=0.6\columnwidth]{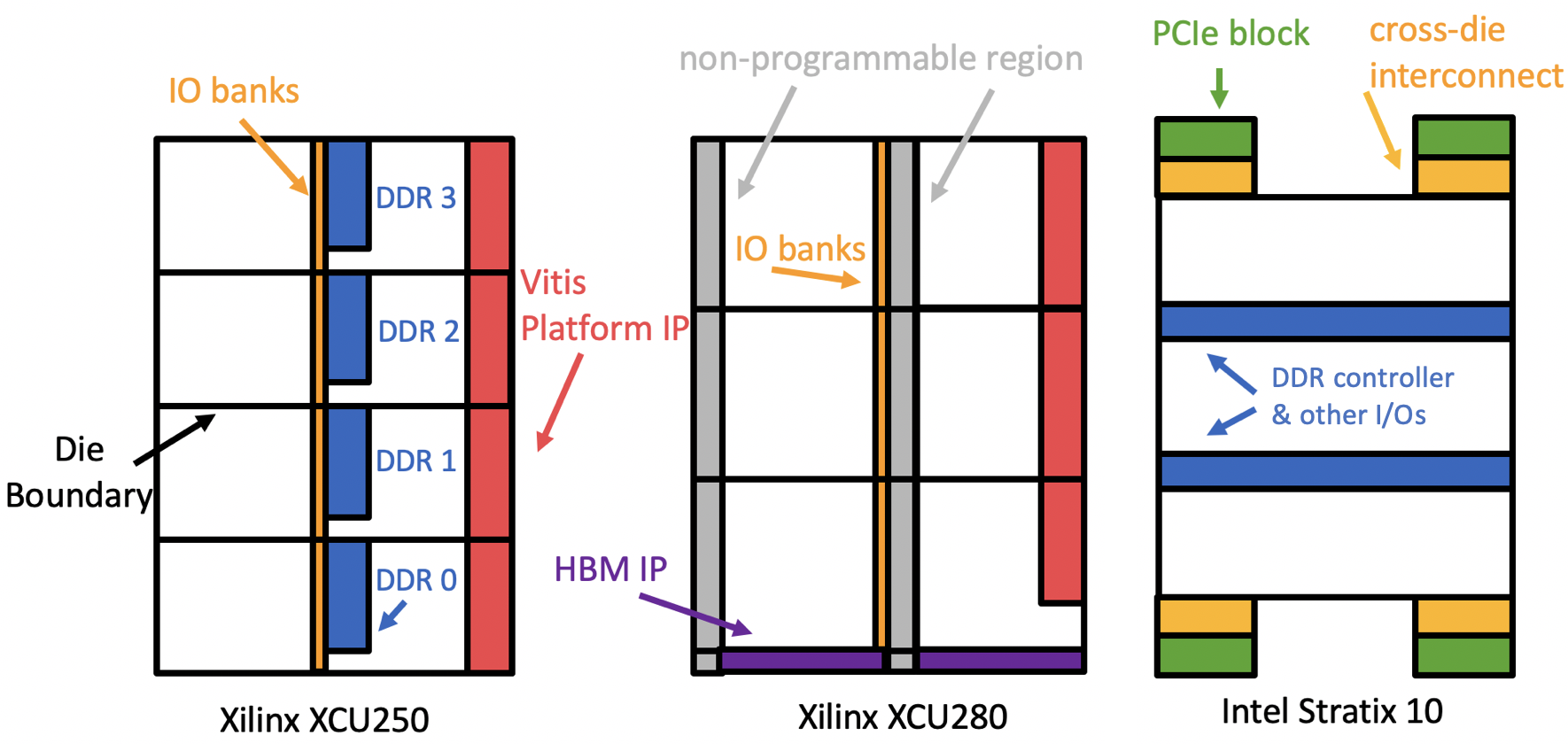}
	\caption{Block diagrams of three representative FPGA architectures: the Xilinx Alveo U250, U280 (based on the Xilinx UltraScale+ architecture), and the Intel Stratix 10.} 
    \label{fig:fpga_arch}
\end{figure}

Figure~\ref{fig:fpga_arch} shows three representative multi-die FPGA architectures, each of which is described in more detail as follows.

\begin{itemize}
    \item The Xilinx Alveo U250 FPGA is one of the largest FPGAs with four dies. All the I/O banks are located in the middle column and the four DDR controller IPs are positioned vertically in a tall-and-slim rectangle in the middle. On the right lies the Vitis platform region~\cite{url-vitis}, which incorporates the DMA IP, the PCIe IP, etc, and serves to communicate with the host CPU. 
    \item The Xilinx Alveo U280 FPGA is integrated with the latest High-Bandwidth Memory (HBM)~\cite{choi2020hls, choi2020hbm, url-hbm}, which exposes 32 independent memory ports at the bottom of the chip. I/O banks are located in the middle columns. Meanwhile, there is a gap region void of programmable logic in the middle.
    \item The Intel Stratix 10 FPGA~\cite{url-stratix10} also sets the DDR controller and I/O banks in the middle of the programmable logic. The embedded multi-die interconnect bridges and the PCIe blocks are distributed at the two sides of the chip, allowing multiple FPGA chips to be integrated together. Although this paper uses the Xilinx FPGAs to demonstrate the idea, our methodology is also applicable to Intel FPGAs and other architectures.
\end{itemize}

Compared to previous generations, the latest multi-die FPGA architectures are divided into disjoint regions, where the region-crossing naturally incurs additional signal delay. In addition, the large pre-located IPs consume significant programmable resources near their fixed locations which may also cause local routing congestion. These characteristics can hamper the existing HLS flows from achieving a high frequency.

\subsection{Motivating Examples}

\begin{figure}[!t]
	\centering
	\includegraphics[width=0.65\columnwidth]{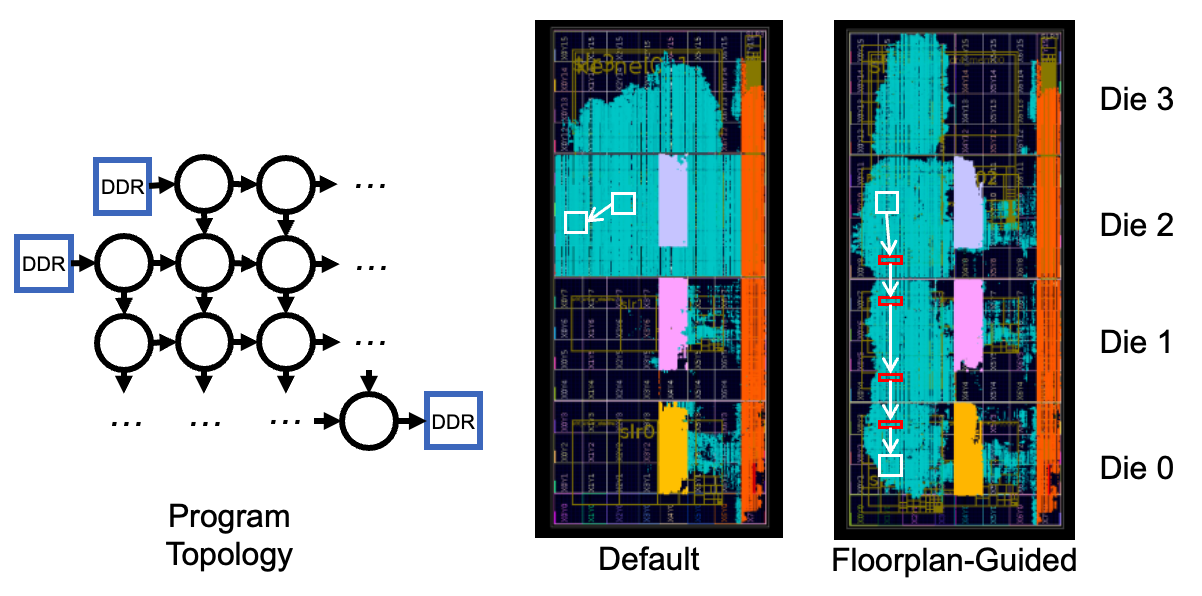}
	\caption{Implementation results of a CNN accelerator on the Xilinx U250 FPGA. Spreading the tasks across the device helps reduce local congestion, while the die-crossing wires are additionally pipelined.}
    \label{fig:polysa_dcp}
\end{figure}

We show two examples to motivate our floorplan-guided HLS approach. First, Figure~\ref{fig:polysa_dcp} shows a CNN accelerator implemented on the Xilinx U250 FPGA. It interacts with three DDR controllers, as marked in grey, pink, and yellow blocks in the figure. In the original implementation result, the whole design is packed close together within die 2 and die 3. To demonstrate our proposed idea, we first manually floorplan the design to distribute the logic in four dies and to avoid overlapping the user logic with DDR controllers. Additionally, we pipeline the FIFO channels connecting modules in different dies as demonstrated in the figure. The manual approach improves the final frequency by 53\%, from 216 MHz to 329 MHz. 

\begin{figure}[!t]
	\centering
	\includegraphics[width=0.6\columnwidth]{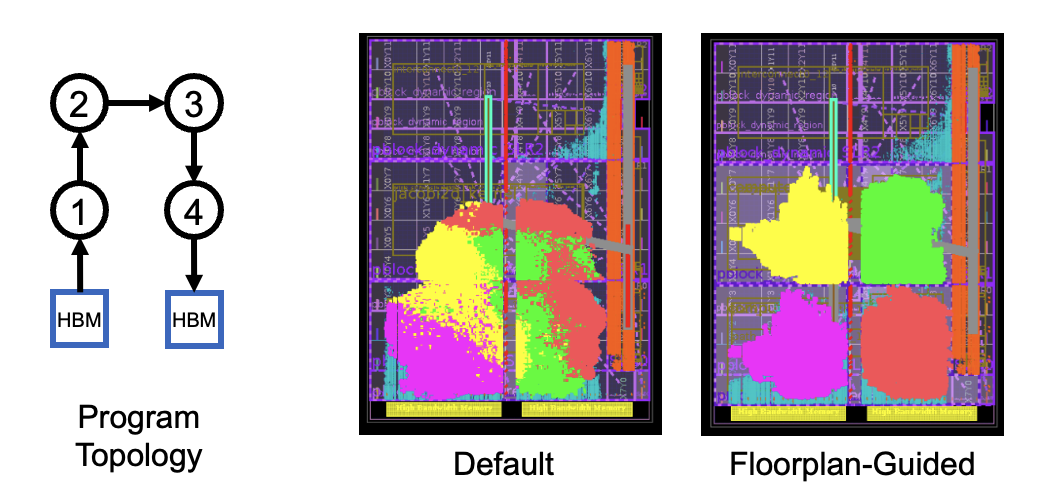}
	\caption{Implementation results of a stencil computing design on U280. Floorplanning during HLS compilation significantly benefits the physical design tools.}
    \label{fig:soda_dcp}
\end{figure}

Second, Figure~\ref{fig:soda_dcp} shows a stencil computation design on the Xilinx U280 FPGA. It consists of four identical tasks in linear topology with each color representing a kernel. In the original implementation, the tool's choice of die-crossing wires is sub-optimal and one kernel may be divided among multiple regions. Instead in our approach, we pre-determine all the die-crossing wires during HLS compilation and pipeline them, so the die boundaries will not cause problems for the placement and routing tool. For this example, we achieve 297 MHz while the design is originally \textit{unroutable}.

\section{Programming Model and Interfaces}
\label{sec:programming-model}

In this section, we present the detailed task-programming model of TAPA and the user interfaces.

\subsection{Basic Concepts}

TAPA dataflow programs explicitly decouple communication and computation. A TAPA program has two types of building blocks: \texttt{streams} and \texttt{tasks}. A {stream} will be mapped to a FIFO in hardware; a task consumes data from input {streams}, performs arbitrary computation, then produces data into other output {streams}. All tasks execute in parallel and communicate with each other through streams.

Listing~\ref{lst:vadd} shows an example TAPA program that instantiates \texttt{PE\_NUM} kernels and each kernel loads two vectors from external memory, adds them up, and stores the results back into external memory. The \texttt{VecAdd} function instantiates the three lower-level tasks and defines the communication streams between them. It takes 3 arguments: 3 \texttt{mmap} interfaces for the 3 vectors and one scalar for the vector length. 3 communication streams are defined in \texttt{VecAdd}. The 3 lower-level tasks are instantiated 4 times in total because there are 2 input vectors, each of which needs its own \texttt{Load}. The \texttt{VecAdd} function is an upper-level task. It is also the top-level task that defines the interface between the kernel and the host. Once the 4 children task instances are instantiated, they will run in parallel and their parent will wait until all children finish.

\begin{listing}[!t]
  \caption{Accelerator task instantiation in TAPA.}
  \label{lst:vadd}
\begin{minted}{cpp}
using namespace tapa;

void Add(istream<int>& a, istream<int>& b, ostream<int>& c, int n) { // to be compiled by Vitis HLS
  for (int i = 0; i < n; ++i) c << (a.read() + b.read()); 
}

void Load(mmap<int> mmap, ostream<int>& stream, int n) { // to be compiled by Vitis HLS
  for (int i = 0; i < n; ++i) stream << mmap[i]; 
} 

void Store(istream<int>& stream, mmap<int> mmap, int n) { // to be compiled by Vitis HLS
  for (int i = 0; i < n; ++i) stream >> mmap[i]; 
}

void VecAdd(mmaps<int, PE_NUM> mem_1, mmaps<int, PE_NUM> mem_2, int n) { // to be compiled by TAPA
  streams<int, PE_NUM, FIFO_DEPTH> str_a, str_b, str_c;
  tapa::task()
      .invoke<PE_NUM>(Load, mem_1, str_a, n)
      .invoke<PE_NUM>(Load, mem_2, str_b, n)
      .invoke<PE_NUM>(Add, str_a, str_b, str_c, n)
      .invoke<PE_NUM>(Store, str_c, mem_2, n);
}
\end{minted}
\end{listing}

\begin{figure}[!t]
	\centering
	\includegraphics[width=0.55\columnwidth]{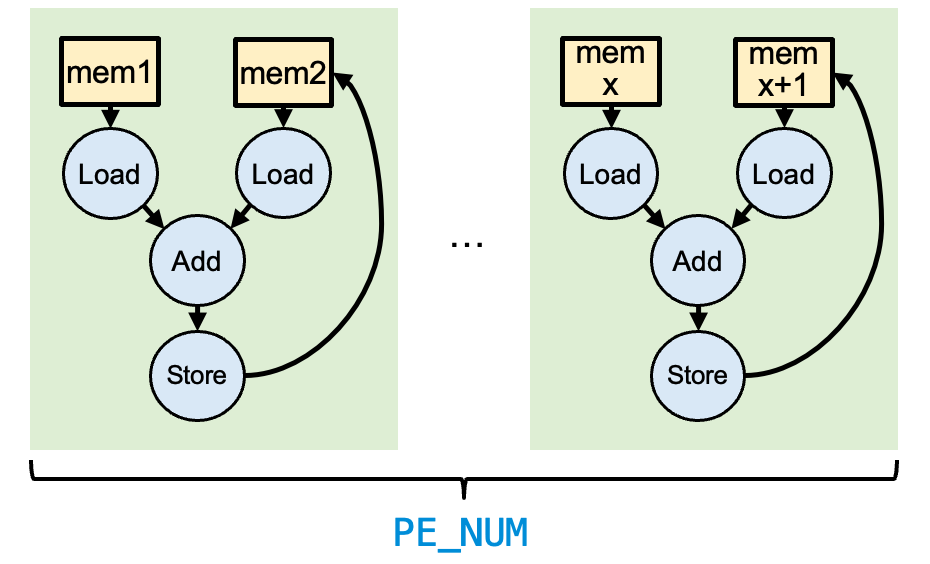}
	\caption{The task graph corresponding to Listing~\ref{lst:vadd}. By adjusting \texttt{PE\_NUM}, we could easily scale out the design for a higher degree of data parallelism.} 
    \label{fig:vadd}
\end{figure}

\subsection{Hierarchical Programming Model}
\label{sec:dfsm-model}

TAPA uses a hierarchical programming model.
Each task is either a leaf that does not instantiate any streams or tasks,
  or a collection of tasks and streams with which the tasks communicate.
A task that instantiates a set of tasks and streams is called the
  \textit{parent task} for that set.
Correspondingly,
  the instantiated tasks are the \textit{children tasks} of their parent,
  which may be parents of their own children.
Each stream must be connected to exactly two tasks.
One of the tasks must act as a \textit{producer} and the other must act as a
  \textit{consumer}. Meanwhile, each task could connect to an arbitrary number of streams.
The producer \textit{streams} tokens to the consumer via the stream
  in the first-in-first-out (FIFO) order.
Each task is implemented as a C++ function,
  which can communicate with each other via the
  \textit{communication interface}.
A parent task instantiates streams and tasks using the
  \textit{instantiation interface}
  and waits until all its child tasks finish.
One of the tasks is designated as the \textit{top-level task},
  which defines the communication interfaces external to the FPGA accelerator,
  i.e., the \textit{system integration interface}.

\subsection{Convenient Programming Interfaces}

\subsubsection{Communication Interface}

TAPA provides separate communication APIs for the producer side and
  the consumer side,
  which use \texttt{ostream} and \texttt{istream} as the interfaces,
  respectively.
The producer of a stream can test the fullness of the stream and append
  tokens to the stream (\texttt{write}) if the stream is not full.
The consumer of a stream can test the emptiness of the stream and remove
  tokens from the stream (destructive \texttt{read}),
  or duplicate the head token without removing it (non-destructive read,
  a.k.a., \texttt{peek}),
  if the stream is not empty.
Read, peek, and write operations can be blocking or non-blocking.

A special token denoting end-of-transaction (EoT) is available to all
  streams.
A process can ``\texttt{close}'' a stream by writing an EoT token to it,
  and a process can ``\texttt{open}'' a stream by reading an EoT token from it.
A process can also test if a stream is closed,
  which is a non-destructive read operation to the stream (\texttt{eot}).
An EoT token does not contain any useful data.
This is designed deliberately to make it possible to break from a pipelined loop
  when an EoT is present,
  for example, in Line~\ref{lln:eot_tapa} of Listing~\ref{lst:ring-tapa-comm}.
Listing~\ref{lst:ring-tapa-comm} shows an example of how the
  communication interfaces are used.

\begin{listing}[!t]
  \caption{Demonstration of TAPA \texttt{peek()} API and the usage of TAPA transactions.}
  \label{lst:ring-tapa-comm}
\begin{minted}{cpp}
void RingNode(istream<Pkt>& node_in,  istream<Pkt>& pe_in,
              ostream<Pkt>& node_out, ostream<Pkt>& pe_out) {
  while (!|\colorpeek{pe\_in.eot()}\label{lln:eot_tapa}|) {
    if (!pe_in.empty()) {
      node_out.write(|\colorread{pe\_in.read()}|);
      if (!node_in.empty() && IsForThisNode(|\colorpeek{node\_in.peek()}|))
        pe_out.write(|\colorread{node\_in.read()}|);
    } else if (!node_in.empty()) {
      Pkt pkt = |\colorread{node\_in.read()}|;
      (IsForThisNode(pkt) ? pe_out : node_out).write(pkt);
    }
  }  // Highlighted are \colorread{destructive read operations} and
}    // \colorpeek{non-destructive read (peek) operations}.
\end{minted}
\end{listing}

\subsubsection{Instantiation Interface}

A parent task can instantiate streams using \texttt{stream<type,capacity>}.
For example,
  \texttt{stream<Pkt,2>} instantiates a stream with capacity 2,
  and data tokens transmitted using this stream have type \texttt{Pkt}.
Tasks are instantiated using \texttt{task::invoke},
  with the first argument being the task function and the rest of the arguments
  being the arguments to the task instance.
This is consistent with \texttt{std::invoke} in the C++ standard library.

\subsubsection{Detached Tasks}
Sometimes making the termination of the program dependent on each kernel function is overkill. For example, a task function may be purely data-driven and we do not have to terminate it on program termination. In that case, TAPA allows users to detach a task on invocation instead of joining it to the parent through the \texttt{task().invoke<detach>()} API. Such tasks can keep running forever as long as input data is available, and the central controller will not check if it is finished or not to determine the termination of the program.

\subsubsection{System Integration Interface}

TAPA uses a unified system integration interface to further reduce programmers'
  burden.
To offload a kernel to an FPGA accelerator,
  programmers only need to call the
  top-level task as a C++ function in the host code.
Since TAPA can extract metadata information, e.g., argument type,
  from the kernel code,
  TAPA will automatically synthesize proper OpenCL host API calls and emit an
  implementation of the top-level task C++ function that can set up the runtime
  environment properly.
As a user of TAPA,
  the programmer can use a single function invocation in the same source code to
  run software simulation, hardware simulation, and on-board execution,
  with the only difference of specifying proper kernel binaries.

\subsection{Asynchronous External Memory Access}
\label{sec:async-mmap}

\begin{figure}[!t]
	\centering
	\includegraphics[width=0.9\columnwidth]{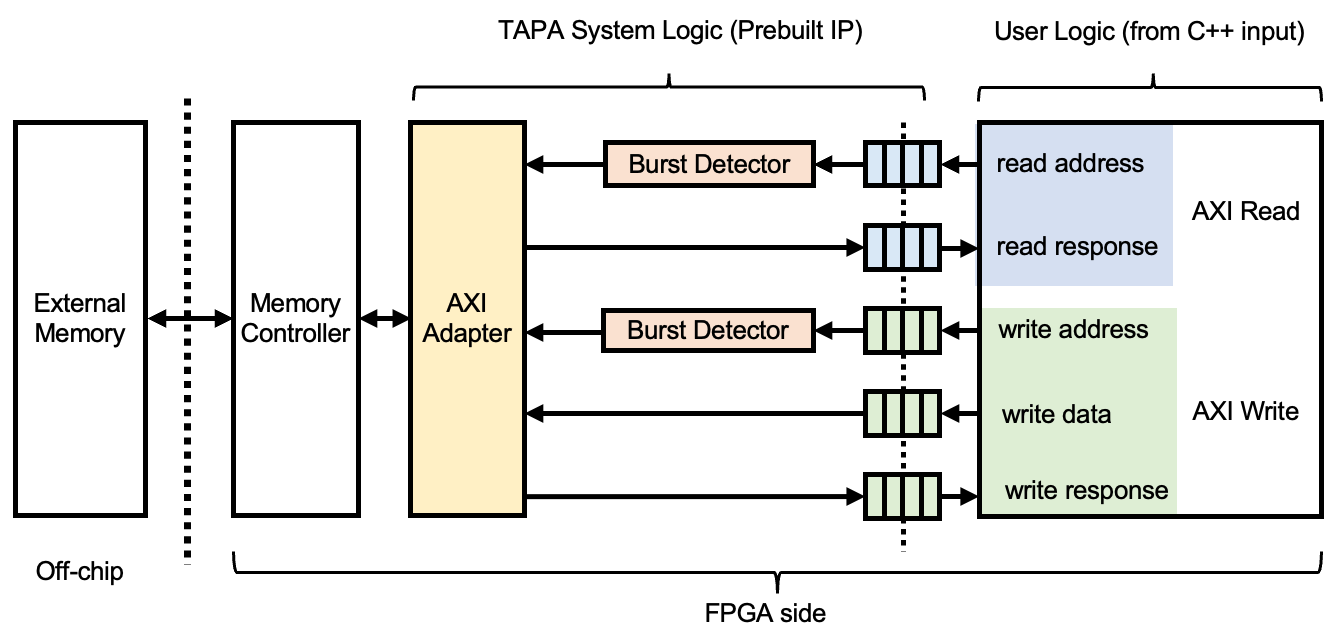}
	\caption{Implementation of the \texttt{async\_mmap} API. The AXI adapter and the burst detector are pre-implemented IPs as part of the TAPA infrastructure. Given a new TAPA program, the RTL generated for the user program will be combined with the TAPA system logic to form a complete working design.}
    \label{fig:async_mmap}
\end{figure}

In order to provide users with more flexibility in accessing the external memory and to reduce the area overhead, we develop the \texttt{async\_mmap} API. The \texttt{async\_mmap} object represents an AXI channel as five independent streams, as shown in Figure~\ref{fig:async_mmap}. To read from an HBM channel, the user only needs to write out the target addresses one by one into the \texttt{read\_addr} stream. The received data will be directly streamed to the user logic through another stream interface. Figure~\ref{lst:async-mmap} shows an example of a sequential read operation. Likewise, to write data out into external memory, the user only needs to push the individual addresses and data to the \texttt{write\_addr} channel. 


Contrary to existing HLS solutions that infer burst transactions at the compile time through static analysis, we implement a runtime burst detection mechanism. The burst detector module will inspect the input addresses and merge sequential reads or writes into a longer burst transaction. Table~\ref{tbl:burst_detector} demonstrates the behavior of the burst detector. The input side is always available to accept new input addresses as long as the downstream AXI adapter is not congested. Internally, the burst detector will keep track of the longest stream of sequential addresses it has seen so far. Once a new input address is not consequential to the previous address, the burst detector will conclude the last transaction and produce a burst transaction to the output. Meanwhile, it will update the starting address of the next burst. In the case that the next input address is not available above a threshold, the burst detector will also conclude and issue out the current burst. The burst detector ensures that memory access will be as efficient as inferring burst transactions statically.

\begin{listing}[!t]
  \caption{The user interface of the \texttt{async\_mmap} API}
  \label{lst:async-mmap}
\begin{minted}{cpp}
template <typename T>
struct async_mmap {
  using addr_t = int64_t; using resp_t = uint8_t;
  tapa::ostream<addr_t>  read_addr;
  tapa::istream<T>       read_data;
  tapa::ostream<addr_t>  write_addr;
  tapa::ostream<T>       write_data;
  tapa::istream<resp_t>  write_resp;
};
\end{minted}
\end{listing}

\begin{listing}[!t]
  \caption{Example of reading data with the \texttt{async\_mmap} API}
  \label{lst:async-mmap-example}
\begin{minted}{cpp}
void load(tapa::async_mmap<Elem>& hbm) {
  for(int i_rd_req = 0, i_rd_resp = 0; i_rd_resp < n; ) { 
#pragma HLS pipeline II=1
    // issue read request
    if (i_rd_req < n && !hbm.read_addr.full()) {
      hbm.read_addr.write(i_rd_req);
      ++i_rd_req;
    }
  
    // receive the read response
    if (!hbm.read_data.empty()) {
      Elem elem = hbm.read_data.read();
      ++i_rd_resp;
    } }
}
\end{minted}
\end{listing}

\begin{table}[]
\caption{The behavior of the burst detector that combines sequential individual memory transactions into burst transactions. In cycles 0-3, four sequential addresses are consumed, and the detector keeps track the length of the current burst. In cycle 4, the new input address (128) is not consecutive to the last input address (64), thus the burst detector put an end to the last burst tracking process and issues a burst transaction on the output side.}
\resizebox{0.7\columnwidth}{!}{
\begin{tabular}{lllllllllll}
\hline
\multicolumn{2}{l}{Cycle}                        & 0  & 1  & 2  & 3  & 4   & 5   & 6   & 7   & ... \\ \hline
Input                           & Read Requests  & 64 & 65 & 66 & 67 & 128 & 129 & 130 & 256 & ... \\ \hline
\multirow{2}{*}{Ouput}          & AXI Read Addr  &    &    &    &    & 64  &     &     & 128 &     \\
                                & AXI Burst Len  &    &    &    &    & 4   &     &     & 3   &     \\ \hline
\multirow{2}{*}{Internal State} & Base Addr      & 64 & 64 & 64 & 64 & 128 & 128 & 128    & 256 &     \\
                                & Length Counter & 1  & 2  & 3  & 4  & 1   & 2   & 3   & 1   &     \\ \hline
\end{tabular}%
}
\label{tbl:burst_detector}
\end{table}

The \texttt{async\_mmap} API gives HLS users a similar level of fine-grained control of the RTL level. Users can use one dedicated task to issue read requests and another task to receive the read responses, which is hardly feasible in the common approach that abstracts the external memory as an array.

\subsection{Overall Compilation Steps}

Figure~\ref{fig:flow} shows the overall compilation flow. Given an input \toolName~program, the tool will compile the top function into RTL and then invoke an existing HLS tool to compile each individual task. For the top function, the tool will analyze and extract how the tasks are interconnected and construct a task graph. The task graph will be used by the floorplan tools to assign each task to one region of the device and determine the pipeline level of each stream. The details of the floorplan process will be discussed in the following sections.

\section{Coupling HLS with Coarse-Grained Floorplanning}
\label{sec:floorplan}

In this section, we present our coarse-grained floorplanning scheme that assigns TAPA tasks to different regions of the programmable fabric. 

Note that the focus of this work is not on improving floorplanning algorithms; instead, we intend to properly use coarse-grained floorplan information to guide HLS and placement.  

\begin{figure}[!t]
	\centering
	\includegraphics[width=0.7\columnwidth]{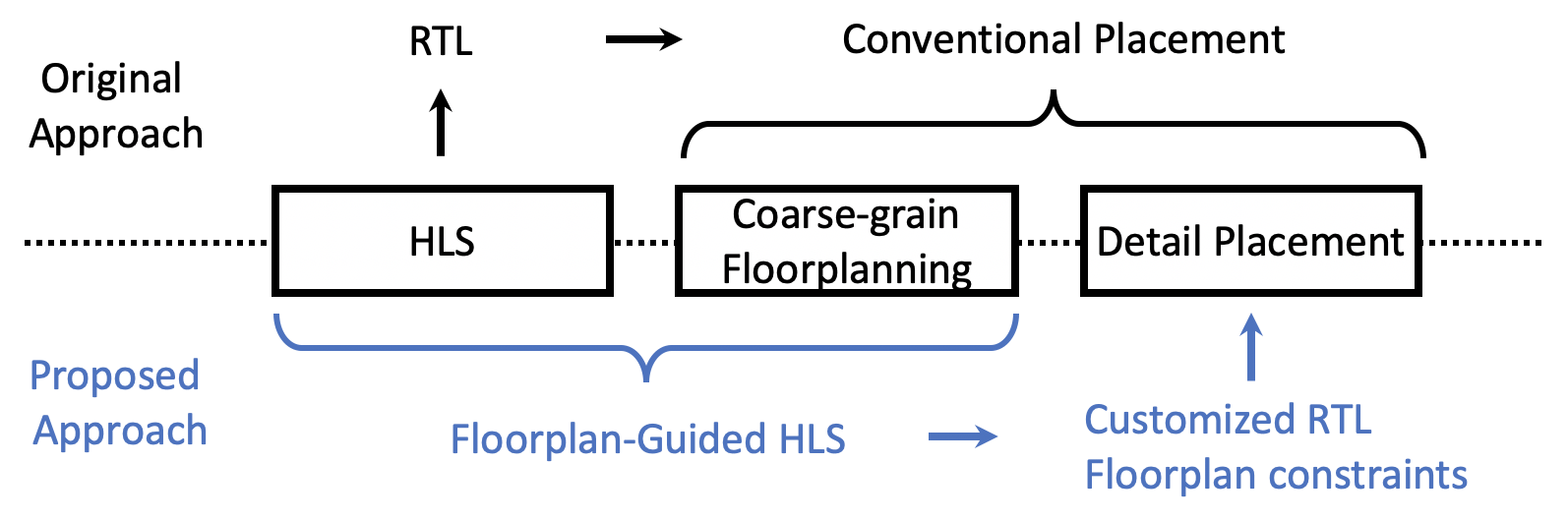}
	\caption{Core Idea of the Proposed Methodology.} 
    \label{fig:core}
\end{figure}

\subsection{Coarse-Grained Floorplanning Scheme}
\label{q2}

Instead of finding a dedicated region with a detailed aspect ratio for each module, we choose to view the FPGA device as a \textbf{\textit{grid}} that is formed by the die boundaries and the large IP blocks. These physical barriers split the programmable fabric apart into a series of disjoint \textbf{\textit{slots}} in the grid where each slot represents a sub-region of the device isolated by die boundaries and IP blocks. Using our coarse-grained floorplanning, we will assign each function of the HLS design to one of these slots. 

For example, for the Xilinx Alveo U250 FPGA, the array of DDR controllers forms a vertical split in the middle column; and there are three horizontal die boundaries. Thus the device can be viewed as a grid of 8 slots in 2 columns and 4 rows. Similarly, the U280 FPGA can be viewed as a grid of 6 slots in 2 columns and 3 rows. 

In this scheme, each slot contains about 700 BRAM\_18Ks, 1500 DSPs, 400K Flip-Flops and 200K LUTs. 
Meanwhile, to reduce the resource contention in each slot, we set a maximum utilization ratio for each slot to guarantee enough blank space. 
Experiments show that such slot sizes are suitable, and HLS has a good handle on the timing quality of the local logic within each slot, as shown in Section~\ref{sec:expriments}. 

\subsection{Problem Formulation}
\label{q3.1}


\textbf{Given}: (1) a graph $G(V, E)$ representing the task-parallel program where $V$ represents the set of tasks and $E$ represents the set of streaming channels between vertices; 
(2) the number of rows $R$ and the number of columns $C$ of the grid representation of the target device; (3) maximum resource utilization ratios for each slot; (4) location constraints such that certain IO modules must be placed nearby certain IP blocks. In addition, we may have constraints that certain vertices must be assigned to the same slot. This is for throughput concerns and  will be explained in Section~\ref{sec:steal}.


\textbf{Goal}:
Assign each $v \in V$ to one of the slots such that (1) the resource utilization ratio\footnote{Based on the estimation of resource utilization by HLS. We can increase the accuracy of the area estimation by optionally running logic synthesis of each task in parallel.} of each slot is below the given limit; (2) the cost function is minimized. We choose the total number of slot-crossings as the cost instead of the total estimated wire lengths. Specifically, the cost function is defined as 
\begin{equation}
\label{eq:objective}
\sum_{e_{ij} \in E} e_{ij}.width \times (|v_i.row - v_j.row| + |v_i.col - v_j.col|)
\end{equation}
where $e_{ij}.width$ is the bitwidth of the FIFO channel connecting $v_i$ and $v_j$ and module $v$ is assigned to the $v.col$-th column and the $v.row$-th row. The physical meaning of the cost function is the sum of the number of slot boundaries that every wire crosses.

\subsection{Solution}
\label{q3.2}

Our problem is relatively small in size as the number of tasks in behavior-level task parallel programs (typically less than thousands) is much smaller than the number of gates in a logic netlist. We adopt the main idea of top-down partitioning-based placement algorithms~\cite{breuer1977class, dunlop1985procedure, maidee2003fast} to solve our problem. Meanwhile, due to the relatively small problem size, we plan to pursue an exact solution for each partitioning process. 

\begin{figure}[!t]
	\centering
	\includegraphics[width=0.75\columnwidth]{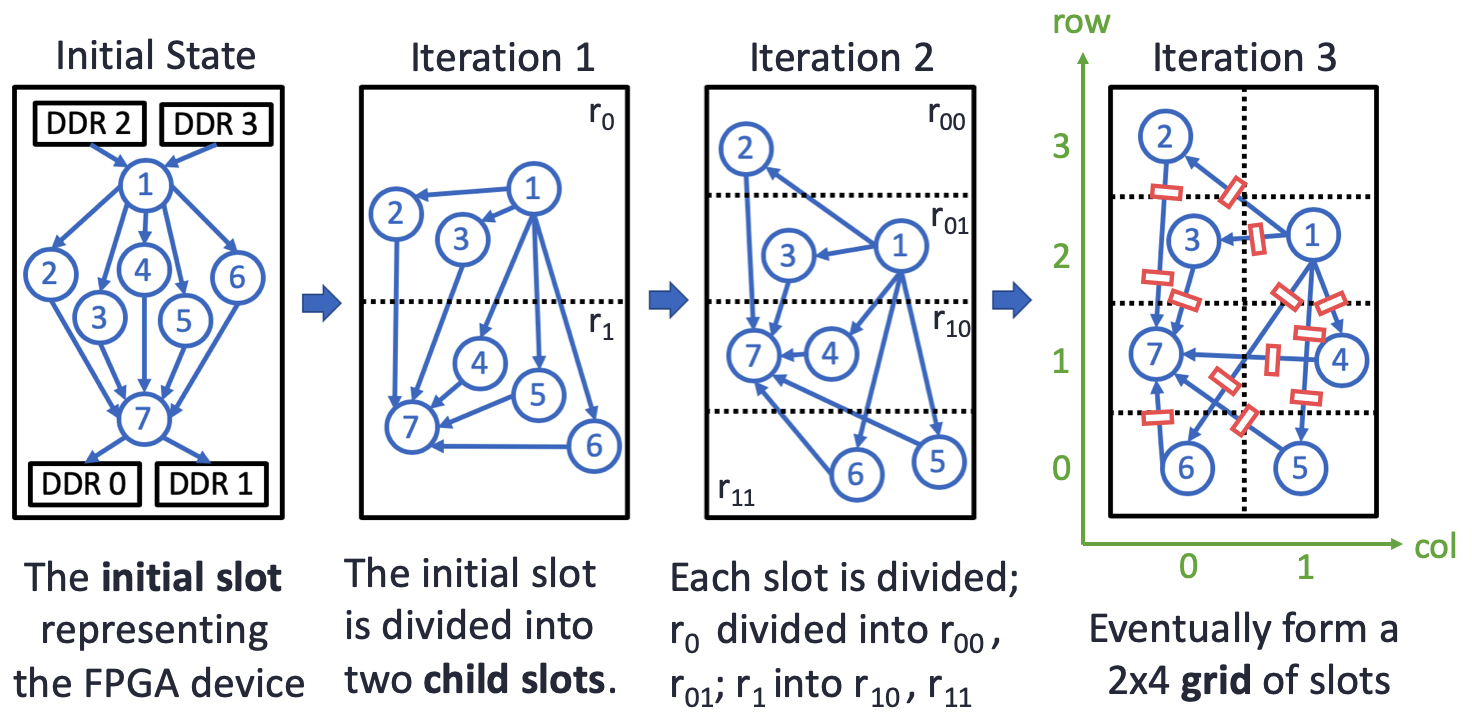}
	\caption{Generating the floorplan for a target $2 \times 4$ grid. Based on the floorplan, all the cross-slot connections will be accordingly pipelined (marked in red) for high frequency.}
    \label{fig:partition}
\end{figure}

Figure~\ref{fig:partition} demonstrates the floorplanning of an example design through three iterations of partitioning. The top-down partitioning-based approach starts with the initial state where all modules are assigned to the same slot, iteratively partitions the current slots in half into two \textbf{\textit{child slots}} and then assigns the modules into the child slots.
Each partitioning involves splitting all of the current slots in half either horizontally or vertically.

We formulate the partitioning process of each iteration using integer linear programming (ILP). In every partitioning iteration, all current slots need to be divided in half. Since some of the modules in a slot may be tightly connected to modules outside of the slot, ignoring such connections can adversely affect the quality of the assignment. Therefore our ILP formulation considers the partitioning of all slots together for an exact solution which is possible due to the small problem size. 
Experiments in Section~\ref{sec:expriments} show that our ILP formulation is solvable within a few seconds or minutes for designs of hundreds of modules.

Performing an N-way partitioning is another potential method. However, compared to our iterative 2-way partitioning, experiments show that it is much slower than iterative 2-way partitioning.

\smallskip
\noindent\textbf{ILP Formulation of One Partitioning Iteration.}

The formulation declares a binary decision variable $v_d$ for each $v$ to denote whether $v$ is assigned to the left or the right child slot during a vertical partitioning (or to the upper or the lower child slot for a horizontal one). 
Let $R$ denote the set of all current slots. For each slot $r \in R$ to be divided, we use $r_v$ to denote the set of all vertices that $r$ is currently accommodating. To ensure that the child slots have enough resources for all modules assigned to them, the ILP formulation imposes the resource constraint for each child slot $r_{child}$ and for each type of on-chip resource.

\begin{equation}
\sum_{v \in r_v}v_d \times v_{area} < (r_{child})_{area}
\end{equation}
where $v_{area}$ is the resource requirement of $v$ and $(r_{sub})_{area}$ represents the available resources in the child slot divided from $r$.

To express the cost function that is based on the coordinates of each module, we first need to express the new coordinates $(v.row, v.col)$ of $v$ based on the previous coordinates $((v.row)_{prev}, (v.col)_{prev})$ and the decision variable $v_d$. For a vertical partitioning, the new coordinates of $v$ will be 
\begin{equation}
v.col = (v.col)_{prev} \times 2 + v_d
\end{equation}
\vspace{-0.5cm}
\begin{equation}
v.row = (v.row)_{prev}
\end{equation}
And for horizontal partitioning, the new coordinates will be 
\begin{equation}
v.row = (v.row)_{prev} \times 2 + v_d
\end{equation}
\vspace{-0.5cm}
\begin{equation}
v.col = (v.col)_{prev}
\end{equation}

Finally, the objective is to minimize the total slot-crossing shown in Formula~(\ref{eq:objective}) for each partitioning iteration.

For the example in Figure~\ref{fig:partition}, Table~\ref{tbl:coordinates} shows the $row$ and $col$ indices of selected vertices in each partitioning iteration. 

\begin{table}[!h]
\caption{Coordinates of selected vertices in Figure~\ref{fig:partition}}
\resizebox{0.7\textwidth}{!}{%
\begin{tabular}{|c|c|c|c|c|}
\hline
       & $v_2$                                                                        & $v_1$                                                                        & $v_4$                                                                       & $v_5$                                                                        \\ \hline \hline
Init   & \multicolumn{4}{c|}{row = 0; col = 0}                                                                                                                                                                                                                                                                                    \\ \hline
iter-1 & \multicolumn{2}{c|}{\begin{tabular}[c]{@{}c@{}}$v_d$ = 1; \\ row = 0 $\times$ 2 + 1 = 1\end{tabular}}                                                       & \multicolumn{2}{c|}{\begin{tabular}[c]{@{}c@{}}$v_d$ = 0;\\ row = 0 $\times$ 2 + 0 = 0\end{tabular}}                                                       \\ \hline
iter-2 & \begin{tabular}[c]{@{}c@{}}$v_d$ = 1; \\ row = 1 $\times$ 2 + 1\end{tabular} & \begin{tabular}[c]{@{}c@{}}$v_d$ = 0; \\ row = 1 $\times$ 2 + 0\end{tabular} & \begin{tabular}[c]{@{}c@{}}$v_d$ = 1;\\ row = 0 $\times$ 2 + 1\end{tabular} & \begin{tabular}[c]{@{}c@{}}$v_d$ = 0; \\ row = 0 $\times$ 2 + 0\end{tabular} \\ \hline
iter-3 & \begin{tabular}[c]{@{}c@{}}$v_d$ = 0; \\ col = 0 $\times$ 2 + 0\end{tabular} & \multicolumn{3}{c|}{\begin{tabular}[c]{@{}c@{}}$v_d$ = 1; \\ col = 0 $\times$ 2 + 1\end{tabular}}                                                                                                                                         \\ \hline
\end{tabular}%
}
\label{tbl:coordinates}
\end{table}




\section{Floorplan-Aware Pipelining}
\label{sec:steal}


Based on the generated floorplan, we aim to \textit{pipeline every cross-slot connection} to facilitate timing closure.
Although HLS has the flexibility to pipeline them to increase the final \textbf{\textit{frequency}}, the additional latency could potentially lead to a large increase of the \textbf{\textit{execution cycles}}, which we need to avoid. This section presents our methods to pipeline slot-crossing connections without hurting the overall throughput of the design.

We will first focus on pipelining the dataflow designs, then extend the method to other types of HLS design. In Section~\ref{sec:s1} we introduce our approach of pipelining with latency balancing; and Section~\ref{sec:s1.2} presents the detailed algorithm. In Sections~\ref{sec:s2} we present how to utilize the internal computation pattern to construct loop-level dataflow graphs that allow more pipelining opportunities.

\subsection{Pipelining Followed by Latency Balancing for Dataflow Designs}
\label{sec:s1}

In our problem, an HLS dataflow design consists of a set of concurrently executed functions communicating through FIFO channels, where each function will be compiled into an RTL module controlled by a finite-state machine (FSM)~\cite{peng1991data}. The rich expressiveness of FSM makes it difficult to statically determine how the additional latency will affect the total execution cycles. Note that our problem is different from other simplified dataflow models such as the Synchronous Data Flow (SDF)~\cite{lee1987synchronous} and the Latency Insensitive Theory (LIT)~\cite{carloni2001theory}, where the firing rate of each vertex is fixed. Unlike SDF and LIT, in our problem, each vertex is an FSM and the firing rate is not fixed and can have complex pattern. 

\begin{figure}[!t]
	\centering
	\includegraphics[width=0.6\columnwidth]{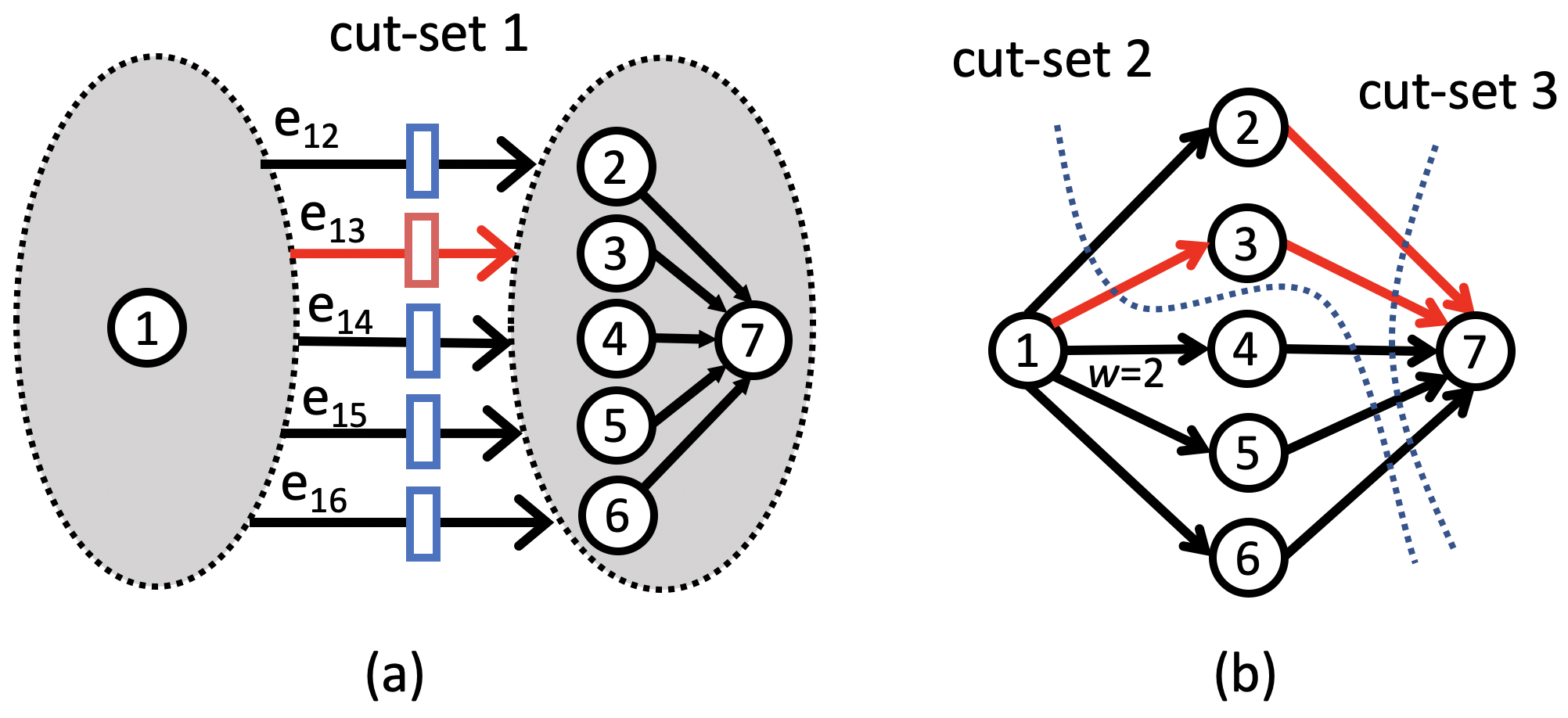}
	\caption{Assume that the edges $e_{13}, e_{37}$ and $e_{27}$ are pipelined according to some floorplan, and each of them carries 1 unit of inserted latency. Also assume that the bitwidth of $e_{14}$ is 2 and all other edges are 1. In the latency balancing step, the optimal solution is adding 2 units of latency to each of $e_{47}, e_{57}, e_{67}$ and 1 unit of latency to $e_{12}$. Note that edge $e_{27}$ and $e_{37}$ can exist in the same cut-set. }
    \label{fig:sdc}
\end{figure}

Therefore, we adopt a conservative approach, where we first pipeline all edges that cross slot boundaries, then balance the latency of parallel paths based on the \textit{cut-set pipelining}~\cite{parhi2007vlsi}. A cut-set is a set of edges that can be removed from the graph to create two disconnected sub-graphs; and if all edges in a cut-set are of the same direction, we could add an equal amount of latency to each edge and the throughput of the design will be unaffected. Figure~\ref{fig:sdc}~(a) illustrates the idea. If we need to add one unit of latency to $e_{13}$ (marked in red) due to the floorplan results, we need to find a cut-set that includes $e_{13}$ and \textit{balance} the latency of all other edges in this cut-set (marked in blue).

Since we can choose different cut-set to balance the same edge, we need to minimize the area overhead. For example, for $e_{13}$, balancing the \texttt{cut-set 2} in Figure~\ref{fig:sdc}~(b) costs smaller area overhead compared to \texttt{cut-set 1} in Figure~\ref{fig:sdc}~(a), as the width of $e_{47}$ is smaller than that of $e_{14}$. Meanwhile, it is possible that multiple edges can be included in the same cut-set. For example, the edges $e_{27}$ and $e_{37}$ are both included in the \texttt{cut-set 3}, so we only need to balance the other edges in \texttt{cut-set 3} once.


Cut-set pipelining is equivalent to balancing the total added latency of every pair of \textit{\textbf{reconvergent paths}}~\cite{parhi2007vlsi}. A path is defined as one or multiple concatenated edges of the same direction; two paths are reconvergent if they have the same source vertex and destination vertex. When there are multiple edges with additional latency from the floorplanning step, we need to find a globally optimal solution that ensures all reconvergent paths have a balanced latency, and the area overhead is minimized.

\subsection{Latency Balancing Algorithm}
\label{sec:s1.2}


\smallskip
\noindent\textbf{Problem Formulation.}

\textbf{Given}: A graph $G(V, E)$ representing a dataflow design that has already been floorplanned and pipelined. Each vertex $v \in V$ represents a function in the dataflow design and each edge $e \in E$ represents the FIFO channel between functions. Each edge $e \in E$ is associated with $e.width$ representing the bitwidth of the edge. For each edge $e$, the constant $e.lat$ represents the additional latency inserted to $e$ in the previous pipelining step.
We use the integer variable $e.balance$ to denote the number of latency added to $e$ in the current latency balancing step. 

\textbf{Goal}: (1) For each edge $e \in E$, compute $e.balance$ such that for any pair of reconvergent paths $\{p_1, p_2\}$, the total latency on each path is the same:
\[\sum_{e \in p_1}{(e.lat + e.balance)} = \sum_{e \in p_2}{(e.lat + e.balance)}\]
and (2) minimize the total area overhead, which is defined as:
\[\sum_{e \in E}{e.balance \times e.width}\]

Note that this problem is different from the classic min-cut problem~\cite{url-min-cut} for DAG. One na\"ive solution is to find a min-cut for every pipelined edge, and increase the latency of the other edges in the cut accordingly. However, this simple method is suboptimal. For example in Figure~\ref{fig:sdc}, since edge $e_{27}$ and $e_{37}$ can be in the same cut-set, we only need to add one unit of latency to the other edges in the cut-set (e.g., $e_{47}$, $e_{57}$ and $e_{67}$) so that all paths are balanced.

\smallskip
\noindent\textbf{Solution.}


We formulate the problem in a restricted form of ILP that can be solved in polynomial time.
For each vertex $v_i$, we associate it with an integer variable $S_i$ that denotes the maximum latency from pipelining between $v_i$ and the sink vertex of the graph. In other words, given two vertices $v_x$ and $v_y$, $(S_x - S_y)$ represents the maximum latency among all paths between the two vertices. Note that we only consider the latency on edges due to pipelining. 

For each edge $e_{ij}$, we have
\begin{equation*}
    S_i \geq S_j + e_{ij}.lat
\end{equation*}

According to our definition, the additional balancing latency added to edge $e_{ij}$ in this step can be expressed as 
\[e_{ij}.balance = (S_i - S_j - e_{ij}.lat)\]
since we want every path from $v_i$ to $v_j$ have the same latency. 

The optimization goal is to minimize the total area overhead, i.e. the weighted sum of the additional depth on each edge:
\begin{equation*}
    minimize \sum_{e_{ij} \in E}{e_{ij}.balance \times e_{ij}.width}
\end{equation*}   

For example, assume that there are two paths from $v_1$ to $v_2$ where path $p_1$ has 3 units of latency from pipelining while $p_2$ has 1 unit. Thus from our formulation, we will select the edge(s) on $p_2$ and add 2 additional units of latency to balance the total latency of $p_1$ and $p_2$ so that the area overhead is minimized.

Our formulation is essentially a system of differential constraints (SDC), in which all constraints are in the form of $x_i - x_j \leq b_{ij}$, where $b_{ij}$ is a constant and $x_i, x_j$ are variables. Because of this restrictive form of constraints, we can solve SDC as a linear programming problem while the solutions are guaranteed to be integers. As a result, it can be solved in polynomial time~\cite{leiserson1991retiming, cong2006efficient}.

If the SDC formulation does not have a solution, there must be a dependency cycle in the dataflow graph~\cite{cong2006efficient}. This means that at least one of the edges in the dependency cycle are pipelined based on the floorplan. In this situation, we will feedback to the floorplanner to constrain those vertices into the same region and then re-generate a new floorplan.

\subsection{Efficient Pipelining Implementation} \label{sec:s2}

\begin{figure}[!t]
	\centering
	\includegraphics[width=0.75\columnwidth]{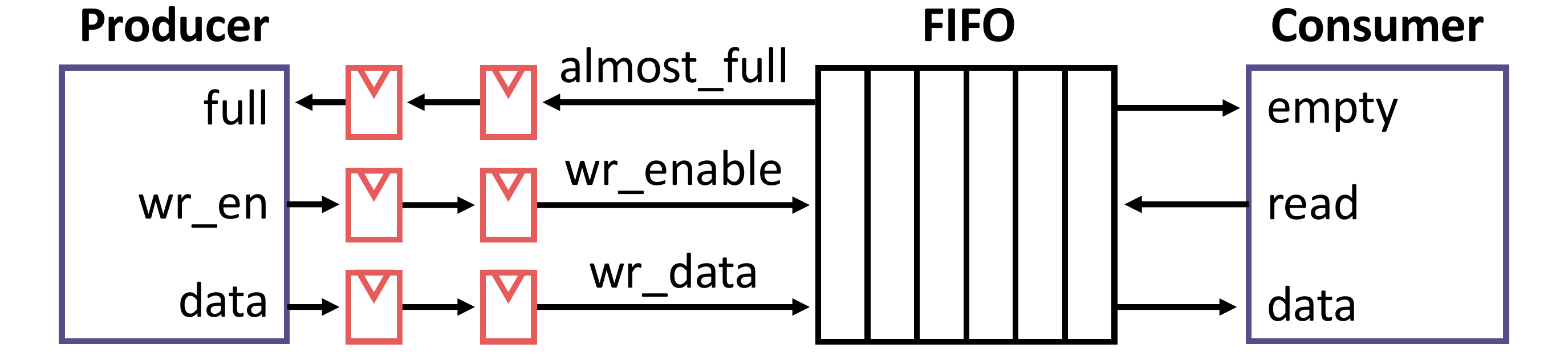}
	\caption{Pipelining FIFO interfaces using almost-full FIFOs.}
    \label{fig:relay_station}
\end{figure}

Figure~\ref{fig:relay_station} shows how we add pipelining to a FIFO-based connection. We adopt FIFOs that assert their \texttt{full} pin before the storage actually runs out, so that we could directly register the interface signals without affecting the functionality.

\section{Optimization for HBM Devices}
\label{sec:opt-hbm}

As will be shown in our evaluation section, the techniques from previous sections will already be effective for a significant timing improvement on DDR-based FPGAs. However, more optimization techniques are needed to squeeze the best performance out of the state-of-the-art HBM-based FPGAs. In this section, we present three major techniques tailored for the unique architecture of HBM-based FPGAs where a large set of independent data channels are clustered closely at the edge of the device. 

\subsection{Reduce BRAM Usage with \texttt{async\_mmap}}

First, we present a system-level optimization to reduce the resource consumption near the HBM blocks by using the \texttt{async\_mmap} API presented in Section~\ref{sec:async-mmap}.
When interacting with the AXI interface, existing HLS tools will buffer the entire burst transactions using on-chip memories. For a 512-bit AXI interface, the AXI buffers generated by Vitis HLS costs 15 BRAM\_18K each for the read channel and the write channel. While this is trivial for conventional DDR-based FPGAs where only a few external DDRs are available, such BRAM overhead becomes a huge problem for HBM devices. To use all 32 HBM channels, the AXI buffers alone take away more than 900 BRAM\_18Ks, which accounts for more than 70\% of the BRAM resources in the bottom SLR. 

However, with the \texttt{async\_mmap} interface, we no longer need to set aside a large buffer to accommodate the data in AXI burst transactions, because the flow control mechanism is explicitly included in the user code (Figure~\ref{lst:async-mmap-example}). Table~\ref{tbl:async_mmap_resource} shows our resource reduction for just one HBM channel.

\begin{table}[]
\caption{Resource comparison of using \texttt{async\_mmap} and using the default array-based abstraction for external memory access. }
\resizebox{0.65\columnwidth}{!}{%
\begin{tabular}{ccccccc}
\hline \textbf{Interface}                             & { \textbf{MHz}} & { \textbf{LUT}} & { \textbf{FF}} & { \textbf{BRAM}} & { \textbf{URAM}} & { \textbf{DSP}} \\ \hline
{ Vitis HLS Default} & 300                                       & 1189                                & { 3740}        & 15                                   & 0                                    & 0                                   \\
{ \texttt{async\_mmap}}                   & 300                                       & 1466                                & 162                                & 0                                    & 0                                    & 0   \\  \hline                              
\end{tabular}%
}
\label{tbl:async_mmap_resource}
\end{table}

\subsection{Automatic HBM Channel Binding}

In the current FPGA HBM architecture, the HBM is divided into 32 channels are physically bundled into eight groups and each group contains four adjacent channels joined by a built-in 4$\times$4 crossbar. The crossbar provides full connectivity within the group. Meanwhile, each AXI interface at the user side can still access any HBM channels outside its group. The data will sequentially traverse through each of the lateral connections until it reaches the crossbar connecting to the target channel, thus inter-group accesses will come with longer latency and potentially less bandwidth due to data link sharing. Therefore, the binding of logical buffers and physical HBM channels will affect the design in two ways:
\begin{itemize}
    \item Since intra-group access is more efficient compared to inter-group accesses, an inferior binding will negatively affect the available bandwidth.
    \item As the HBM channels are hardened to fixed locations, the binding also affects the placement and routing of the logic that connect to HBM. Thus an unoptimized binding may cause local congestion in the programmable logic nearby the HBM channels.
\end{itemize}
Existing CAD tools require that users explicitly specify the mapping of all HBM channels, which requires users to master low-level architecture details. Also, since the binding does not affect the correctness of the design, users are often unaware of suboptimal choices. 

To alleviate the problem, we propose a semi-automated solution. We observe that very often the design only involves intra-group HBM accesses. In this case, the binding decision does not affect the HBM bandwidth and latency and only impacts the placement and routing of nearby logic. Therefore, we implement an API where users could specify the partial binding of channels, or none at all if desired, and let TAPA automatically determines the binding for the rest. 

Specifically, we incorporate the HBM binding process into our floorplanning step. We treat the number of available HBM channels as another type of resource for the slots. Therefore, slots that are directly adjacent to HBM blocks will be treated as having the corresponding number of HBM channels, while other slots will have zero available HBM channel resources. Meanwhile, each task that directly interacts with the HBM channel is treated as requiring one unit of HBM channel resources, and other tasks will be regarded as not requiring HBM resources.

\subsection{Generating Multiple Floorplan Candidates}

By default, TAPA will only generate one floorplan solution where we will prioritize a balanced distribution of logic and then accordingly pipeline the inter-slot connections. However, due to the severe local congestion around the bottom die in an HBM device, we need to explore the different trade-offs between logic resource usage and routing resource usage, especially die-crossing wires. One floorplan solution may use fewer logic resources in the bottom die but requires more die-crossing wires as logic are pushed to the upper regions; another solution may have the opposite effect. We observe that very often it is unpredictable which factor is more important for a given design until the routing process is done. Note that each different floorplan solution comes with corresponding pipelining schemes that best suit the floorplan results.

Instead of generating only one floorplan solution, we can generate a set of Pareto-optimal points and run physical design concurrently to explore the best results. In our formulation of the floorplan problem, we have a parameter to control the maximal logic resource utilization of each island. Reducing this parameter will reduce local logic resource usage and increase global routing resource usage and vice versa. Therefore, we sweep through a range of this parameter to generate a set of slightly different floorplans and implement them in parallel to achieve the highest frequency.

\section{Experiments}
\label{sec:expriments}

\subsection{Implementation Details}

TAPA is implemented in C++ and Python. We implement our prototype to interface with the CAD flow for AMD/Xilinx FPGAs, including Vitis HLS, Vivado and Vitis (2021.2). 
We use the Python MIP package~\cite{url-mip} coupled with Gurobi~\cite{url-gurobi} to solve the various ILP problems introduced in previous sections. We generate \texttt{tcl} constraint files to be used by Vivado to enforce our high-level floorplanning scheme. 

Meanwhile, we turn off the hierarchy rebuild process during RTL synthesis~\cite{url-vivado} to prevent the RTL synthesis tool from introducing additional wire connections between RTL modules. The hierarchy rebuild step first flattens the hierarchy of the RTL design and then tries to rebuild the hierarchy. As a result, hierarchy rebuild may create unpredictable new connections between modules. As a result, if two modules are floorplanned far apart, these additional wires introduced during RTL synthesis will be under-pipelined as they are unseen during HLS compilation. Note that disabling this feature may lead to slight differences in the final resource utilization.

We test out designs on the Xilinx Alveo U250 FPGA\footnote{The U250 FPGA contains 5376 BRAM18K, 12288 DSP48E, 3456K FF and 1728K LUT} with 4 DRAMs and the Xilinx Alveo U280 FPGA\footnote{The U280 FPGA contains 4032 BRAM18K, 9024 DSP48E, 2607K FF and 434K LUT} with High-Bandwidth Memory (HBM). As the DDR controllers are distributed in the middle vertical column while the HBM controller lies at the bottom row, these two FPGA architectures present different challenges to the CAD tools. Thus it is worthwhile to test them separately. 

To run our framework, users first specify how they want to divide the device. By default, we divide the U250 FPGA into a 2-column $\times$ 4-row grid and the U280 FPGA into a 2-column $\times$ 3-row grid, matching the block diagram of these two architectures shown in Figure~\ref{fig:fpga_arch}.
To control the floorplanning, users can specify the maximum resource utilization ratio of each slot. The resource utilization is based on the estimation by HLS.
Users can also specify how many levels of pipelining to add based on the number of boundary crossings. By default, for each boundary crossing we add two levels of pipelining to the connection. The processed design is integrated with the Xilinx Vitis infrastructure to communicate with the host.

\subsection{Benchmarks}

\begin{figure}[!t]
	\centering
	\includegraphics[width=0.6\columnwidth]{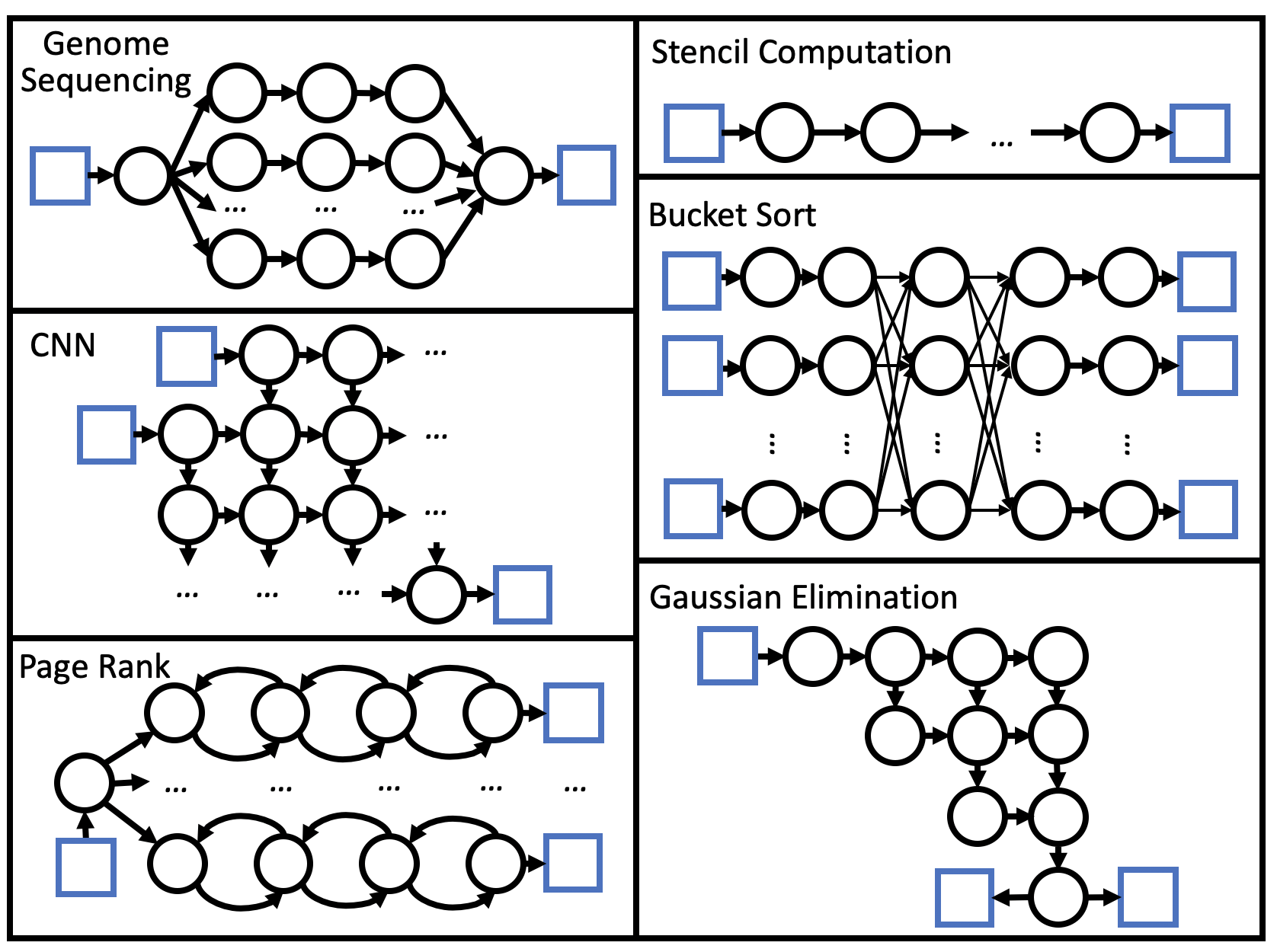}
	\caption{Topologies of the benchmarks. Blue rectangles represent external memory ports and black circles represent computation kernels of the design. In the genome sequencing design, the arrows represent BRAM channels; in other designs, the arrows represent FIFO channels.}
    \label{fig:topology}
\end{figure}


We use two groups of benchmarks to demonstrate the proposed methodologies. We first include six benchmarks that are originally used in AutoBridge~\cite{fpga21-autobridge} to showcase the frequency improvement from co-optimization of HLS and physical design. AutoBridge uses six representative benchmark designs with different topologies and changes the parameter of the benchmarks to generate a set of designs with varying sizes on both the U250 and the U280 board. The six designs are all large-scale designs implemented and optimized by HLS experts.  Figure~\ref{fig:topology} shows the topology of the benchmarks. Note that even for those benchmarks that seem regular (e.g., CNN), the location constraints from peripheral IPs can highly distort their physical layouts.

\begin{itemize}
    \item The stencil designs created by the SODA~\cite{chi2018soda} compiler are a set of kernels in linear topologies.
    \item The genome sequencing design~\cite{guo2019hardware} performing the Minimap2 overlapping algorithm~\cite{li2018minimap2} has processing elements (PE) in broadcast topology. This benchmark is based on shared-memory communication and all other benchmarks are dataflow designs.
    \item The CNN accelerators created by the PolySA~\cite{cong2018polysa} compiler are in a grid topology.
    \item The HBM graph processing design~\cite{chi2020extending} performs the page rank algorithm. It features eight sets of processing units and one central controller. This design also contains dependency cycles, if viewed at the granularity of computing kernels.
    \item The HBM bucket sort design adapted from ~\cite{samardzic2020bonsai} which includes 8 parallel processing lanes and two fully-connected layers.
    \item The Gaussian elimination designs created by AutoSA~\cite{wang2021autosa} are in triangle topologies. 
\end{itemize}

In addition, we include three additional benchmarks that use a large number of HBM channels to demonstrate the newly-added HBM-specific optimizations. All of the three additional benchmarks will still fail to route with the original AutoBridge. However, our latest optimizations enable them to route successfully with high frequencies.

\begin{itemize}

    \item The Scalable and Automatic Stencil Acceleration Framework (SASA)~\cite{sasa} accelerators where one version uses 24 channels, and the other one uses 27 channels. Compared to the SODA stencil accelerator used in the original AutoBridge paper, the SASA accelerator also has a much more complicated topology.
    \item The HBM sparse matrix-matrix multiply (SpMM) accelerator~\cite{song2022sextans} that uses 29 HBM channels. 
    \item The Sparse matrix-vector multiply (SpMV) accelerators~\cite{song2022serpens} where one version uses 20 HBM channels, and another version uses 28 HBM channels.

\end{itemize}

\subsection{Original Evaluation of AutoBridge}
By varying the size of the benchmarks, in total we have tested the implementation of 43 designs with different configurations. Among them, 16 designs failed in routing or placement with the baseline CAD flow, compared AutoBridge which succeeds in routing all of them and achieves an average of 274 MHz. For the other 27 designs, we improve the final frequency from 234 MHz to 311 MHz on average.  
In general, we find that AutoBridge is effective for designs that use up to about 75\% of the available resources. We execute our framework on an Intel Xeon CPU running at 2.2GHz. Both the baseline designs and optimized ones are implemented using Vivado with the highest optimization level.  The final design checkpoint files of all experiments are available in our open-sourced repository. 

In some experiments, we may find that the optimized versions have even slightly smaller resource consumption. Possible reasons are that we adopt a different FIFO template and disable the hierarchy rebuild step during RTL synthesis. Also, as the optimization leads to very different placement results compared to those of the original version, we expect different optimization strategies will be adopted by the physical design tools. The correctness of the code is verified by cycle-accurate simulation and on-board execution.  

Next, we present the detailed results of each benchmark. 

\smallskip
\noindent\textbf{SODA Stencil Computation.}

\begin{figure}[!h]
\centering
\includegraphics[width=0.8\columnwidth]{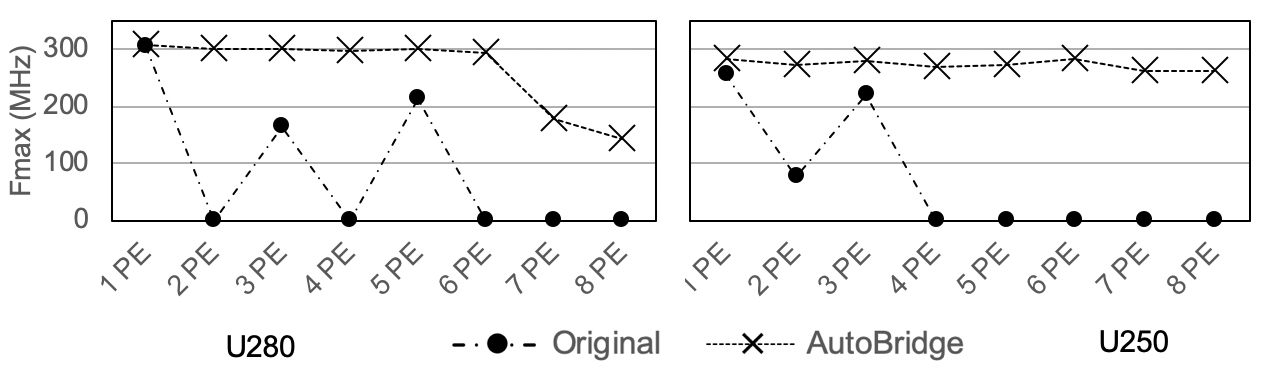}
\caption{Results of the SODA stencil computation designs.  Zero Fmax represents placement or routing failures.}
\label{fig:soda_u280}
\end{figure}

For the stencil computing design, the kernels are connected in a chain format through FIFO channels. By adjusting the number of kernels, we can vary the total size of the design. We test anywhere from one kernel up to eight kernels, and Figure~\ref{fig:soda_u280} shows final frequency of the eight design configurations on both U250 and U280 FPGAs. In the original flow, many design configurations fail in routing due to routing resource conflicts. Those that are routed successfully still achieve relatively low frequencies. In comparison, with the help of AutoBridge, all design configurations are routed successfully. On average, we improve the timing from 86 MHz to 266 MHz on the U280 FPGA, and from 69 MHz to 273 MHz on the U250 FPGA.

Starting from the seven-kernel design, we observe a frequency decrease on the U280 FPGA. This is because each kernel of the design is very large and uses about half the resources of a slot; thus starting from the seven-kernel design on the relatively small U280, two kernels have to be squeezed into one slot which will cause more severe local routing congestion. Based on this phenomenon, we recommend that users avoid designing very large kernels and instead split the functionality into multiple functions to allow the tool more flexibility in floorplanning the design.
   
\smallskip
\noindent\textbf{CNN Accelerator.}

\begin{figure}[!h]
	\centering
	\includegraphics[width=0.8\columnwidth]{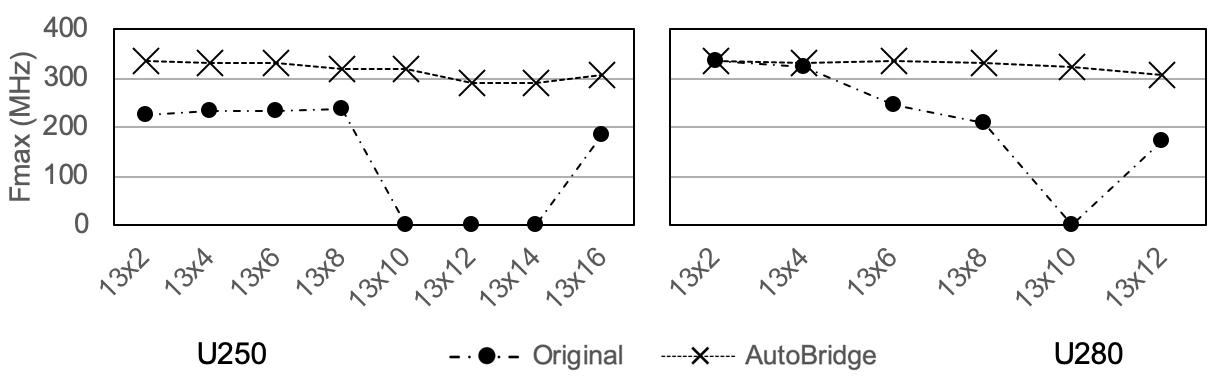}
	\caption{Results of the CNN accelerator designs.}
    \label{fig:polysa_curve}
\end{figure}

The CNN accelerator consists of identical PEs in a regular grid topology. We adjust the size of the grid from a $2 \times 13$ array up to a $16 \times 13$ array to test the robustness of AutoBridge. Figure~\ref{fig:polysa_curve} shows the result on both U250 and U280 FPGAs. 

Although the regular 2-dimensional grid structure is presumed to be FPGA friendly, the actual implementation results from the original tool flow is not satisfying. With the original tool flow, even small-size designs are bounded at around 220 MHz when targeting U250. Designs of larger sizes will fail in placement ($13 \times 12$) or routing ($13 \times 10$ and $13 \times 14$). Although the final frequency is high when the design is small for the original tool flow targeting U280,  the timing quality is steadily dropping as the designs become larger. 

In contrast, AutoBridge improves from 140 MHz to 316 MHz on U250 on average, and from 214 MHz to 328 MHz on U280.
Table~\ref{tbl:polysa} lists the resource consumption and cycle counts of the experiments on U250. Statistics on U280 are similar and are omitted here. 

\begin{table}[!h]
\caption{Post-placement results of the CNN designs on U250. The design point of $13 \times 12$ failed placement and $13 \times 10$ and $13 \times 14$ failed routing with the original tool flow.}
\resizebox{0.8\textwidth}{!}{%
\setlength\tabcolsep{4pt}
\begin{tabular}{ccccccccccc}
\hline
\multirow{2}{*}{Size} & \multicolumn{2}{c}{LUT(\%)} & \multicolumn{2}{c}{FF(\%)} & \multicolumn{2}{c}{BRAM(\%)} & \multicolumn{2}{c}{DSP(\%)} & \multicolumn{2}{c}{Cycle} \\ \cline{2-11} 
                      & orig         & opt          & orig         & opt         & orig          & opt          & orig         & opt          & orig        & opt         \\ \hline
13x2                  & 17.82        & 17.90        & 14.11        & 14.25       & 21.69         & 21.67        & 8.57         & 8.57         & 53591       & 53601       \\
13x4                  & 23.52        & 23.59        & 18.98        & 19.04       & 25.74         & 25.73        & 17.03        & 17.03        & 68630       & 68640       \\
13x6                  & 29.26        & 29.24        & 23.86        & 23.80       & 29.80         & 29.78        & 25.50        & 25.50        & 86238       & 86248       \\
13x8                  & 34.98        & 34.90        & 28.72        & 28.56       & 33.85         & 33.84        & 33.96        & 33.96        & 103882      & 103892      \\
13x10                 & 40.71        & 40.48        & 33.58        & 33.25       & 37.91         & 37.89        & 42.42        & 42.42        & 121472      & 121491      \\
13x12                 & -            & 46.18        & -            & 38.06       & -             & 41.95        & -            & 50.89        & 139098      & 139108      \\
13x14                 & 52.10        & 51.92        & 43.28        & 42.93       & 46.02         & 46.00        & 59.35        & 59.35        & 156715      & 156725      \\
13x16                 & 57.82        & 57.61        & 48.13        & 47.70       & 50.07         & 50.06        & 67.81        & 67.81        & 174377      & 174396      \\ \hline
\end{tabular}%
}
\label{tbl:polysa}
\end{table}

\smallskip
\noindent\textbf{Gaussian elimination.}

The PEs in this design form a triangle topology. We adjust the size of the triangle and test on both U250 and U280. Table~\ref{tbl:lu} shows the results. On average, we improve the frequency from 245 MHz to 334 MHz on U250, and from 223 MHz to 335 MHz on U280.

\begin{figure}[!h]
	\centering
	\includegraphics[width=0.8\columnwidth]{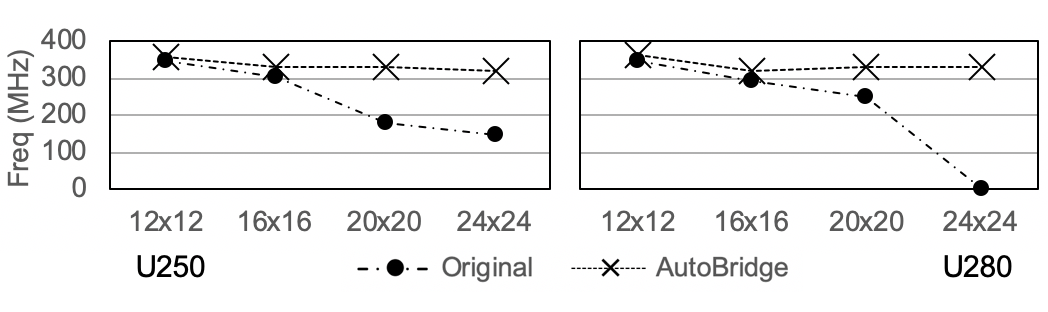}
	\caption{Results of the Gaussian elimination designs.}
    \label{fig:lu_curve}
\end{figure}

\begin{table}[!h]
\caption{Results of Gaussian Elimination Designs on U250.}
\resizebox{0.8\textwidth}{!}{%
\setlength\tabcolsep{4pt}

\begin{tabular}{cccccccccccccc}
\hline
\multirow{2}{*}{Size} & \multicolumn{2}{c}{LUT(\%)} & \multicolumn{2}{c}{FF(\%)} & \multicolumn{2}{c}{BRAM(\%)}      & \multicolumn{2}{c}{DSP(\%)} & \multicolumn{2}{c}{Cycle} \\ \cline{2-11} 
                      & orig         & opt          & orig         & opt         & orig  & opt                       & orig         & opt          & orig          & opt           \\ \hline
12$\times$12          & 18.58        & 18.69        & 13.05        & 13.14       & 13.24 & 13.21 & 2.79         & 2.79         & 758           & 781           \\
16$\times$16          & 26.62        & 26.68        & 17.36        & 17.30       & 13.24 & 13.21                     & 4.99         & 4.99         & 1186          & 1209          \\
20$\times$20          & 38.55        & 38.28        & 23.46        & 23.38       & 13.24 & 13.21                     & 7.84         & 7.84         & 1728          & 1738          \\
24$\times$24          & 54.05        & 53.59        & 32.16        & 32.06       & 13.24 & 13.21                     & 11.34        & 11.34        & 2361          & 2375         \\ \hline
\end{tabular}%
}
\label{tbl:lu}
\end{table}

\smallskip
\noindent\textbf{HBM Bucket Sort.}

The bucket sort design has two complex fully-connected layers. Each fully-connected layer involves an $8 \times 8$ crossbar of FIFO channels, with each FIFO channel being 256-bit wide. AutoBridge pipelines the FIFO channels to alleviate the routing congestion. Table~\ref{tbl:bucket-sort} shows the frequency gain, where we improve from 255 MHz to 320 MHz on U280. As the design requires 16 external memory ports and U250 only has 4 available, the test for this design is limited to U280 only.

Because the original source code has enforced a BRAM-based implementation for some small FIFOs, which results in wasted BRAM resources, the results of AutoBridge have slightly lower BRAM and flip-flop consumption than the original implementation. In comparison, we use a different FIFO template that chooses the implementation style (BRAM-based or shift-register-based) based on the area of the FIFO. Cycle accurate simulation has proven the correct functionality of our optimized implementation.

\begin{table}[!h]
\caption{Results of the Bucket Sort Design on U280.}
\resizebox{0.7\columnwidth}{!}{%
\begin{tabular}{ccccccc}
\hline
\multicolumn{1}{c}{} & \multicolumn{1}{c}{Fmax (MHz)} & \multicolumn{1}{c}{LUT \%} & \multicolumn{1}{c}{FF \%} & \multicolumn{1}{c}{BRAM \%}  & \multicolumn{1}{c}{DSP \%} & \multicolumn{1}{c}{Cycle} \\ \hline
Original    & 255 & 28.44  & 19.11 & 16.47 & 0.04 &  78629 \\
Optimized     & 320   & 29.39    & 16.66  & 13.69 & 0.04 & 78632     \\ \hline
\end{tabular}%
}
\label{tbl:bucket-sort}
\end{table}

\smallskip
\noindent\textbf{HBM Page Rank.}

This design incorporates eight sets of processing units, each interfacing with two HBM ports. There are also centralized control units that exchange control information with five HBM ports. 
Table~\ref{tbl:pagerank} shows the experiment results and we improve the final frequency from 136 MHz to 210 MHz on U280. 

\begin{table}[!h]
\caption{Results of the Graph Processing Design on U280.}
\resizebox{0.7\columnwidth}{!}{%
\begin{tabular}{ccccccc}
\hline
\multicolumn{1}{c}{} & \multicolumn{1}{c}{Fmax (MHz)} & \multicolumn{1}{c}{LUT \%} & \multicolumn{1}{c}{FF \%} & \multicolumn{1}{c}{BRAM \%}  & \multicolumn{1}{c}{DSP \%} & \multicolumn{1}{c}{Cycle} \\ \hline
Original    & 136 & 38.56  & 26.97 & {\color[HTML]{333333} 26.74} & 14.43 & 120458  \\
Optimized    & 210   & 39.49    & 27.53  & {\color[HTML]{333333} 30.08} & 14.43 & 120495     \\ \hline
\end{tabular}%
}
\label{tbl:pagerank}
\end{table}

\subsection{HBM-Specific Optimizations}
\label{sec:hbm-expr}

In this section, we use four real-world designs from premium academic conferences or journals to demonstrate the effects of our HBM-specific optimizations. We select those designs as they use a large number of HBM channels, which brings about a serious timing closure challenge. 

\smallskip
\noindent\textbf{HBM SpMM and SpMV Accelerators.}

The sparse matrix-matrix multiplication (SpMM) and the sparse matrix-vector multiplication (SpMV) accelerators leverage the {\tt async\_mmap} API, automatic HBM channel binding, die-crossing wire adjusting, and multi-floorplan generation to achieve the best performance. We implement two versions of the SpMV accelerator, SpMV\_A24 and SpMV\_A16, with different numbers of parallel processing elements.
We report the user clock and HBM clock frequencies and the resource utilization 
in Table~\ref{tbl:spmmmv}. We have improved both the user clock and the HBM clock frequencies 
for the three designs. Especially for SpMV\_A24, we have improved
the user clock frequency from
193 MHz to 283 MHz and the HBM clock frequency from
430 MHz to 450 MHz. With the {\tt async\_mmap}, we significantly
reduced BRAM utilization -- for SpMM and SpMV\_A24, 
we reduced 10\% of the total BRAM utilization.

\begin{table}[!h]
\caption{Frequency Improvement and Area Reduction of SpMM and SpMV on U280.}
\resizebox{0.8\columnwidth}{!}{%
\begin{tabular}{ccccccc}
\hline
\multicolumn{1}{c}{} & \multicolumn{1}{c}{Fuser/Fhbm (MHz)} & \multicolumn{1}{c}{LUT \%} & \multicolumn{1}{c}{FF \%} & \multicolumn{1}{c}{BRAM \%} & \multicolumn{1}{c}{URAM \%}  & \multicolumn{1}{c}{DSP \%}  \\ \hline
Orig, SpMM    & 216/424  & 37.35  & 31.34  & 71.55  & 53.33  & 40.97 \\
Opt, SpMM    & 245/427   & 39.40  & 28.47  & 60.91  & 53.33 & 40.97     \\ \hline
Orig, SpMV\_A16   & 233/450  & 21.70  & 17.36  & 42.63  & 40.00  & 8.89 \\
Opt, SpMV\_A16    & 299/450  & 22.75  & 15.06  & 35.52  & 40.00  & 8.89     \\ \hline
Orig, SpMV\_A24   & 193/430  & 27.95  & 22.46  & 58.31  & 40.00  & 14.92 \\
Opt, SpMV\_A24    & 283/450  & 29.76  & 19.11  & 48.21  & 40.00  & 14.92    \\ \hline
\end{tabular}%
}
\label{tbl:spmmmv}
\end{table}

\smallskip
\noindent\textbf{HBM Stencil Accelerators by SASA.}

The SASA design incorporates the {\tt async\_mmap} API, automatic HBM
channel binding, die-crossing wire adjusting, and floorplan candidate generation to push the user clock frequency above 225 MHz and the HBM clock frequency to 450 MHz, which enables the accelerator to fully utilize the HBM bandwidth. For stencil algorithms that have a low number of iterations, SASA will leverage efficient spatial parallelism where each kernel read one tile of input data and additional halo data from neighboring tiles at the start. Then each kernel performs the computation for all iterations (if any) without synchronization. Each kernel works in a streaming pattern and uses at least two HBM banks to store the input and output. The original design based on \texttt{mmap} fails to meet the frequency requirement. With the {\tt async\_mmap} API, we are able to significantly reduce the BRAM utilization. With all optimizations, the two selected designs achieve 241 MHz and 250 MHz respectively.

\begin{table}[!h]
\caption{Frequency Improvement and Area Reduction of SASA on U280.}
\resizebox{0.8\columnwidth}{!}{%
\begin{tabular}{ccccccc}
\hline
\multicolumn{1}{c}{} & \multicolumn{1}{c}{Fuser/Fhbm (MHz)} & \multicolumn{1}{c}{LUT \%} & \multicolumn{1}{c}{FF \%} & \multicolumn{1}{c}{BRAM \%} & \multicolumn{1}{c}{URAM \%} & \multicolumn{1}{c}{DSP \%}  \\ \hline
Orig, SASA-1    & 169/295  & 32.2  & 21.98  & 18.45  & 0 & 17.02  \\ 
Opt, SASA-1    & 241/450   & 36.04  & 18.09 & 0  & 0 & 17.02     \\ \hline
Orig, SASA-2    & Failed/Failed  & 35.9  & 30.94  & 23.88  & 0 & 47.87 \\ 
Opt, SASA-2    & 250/450   & 40.63  & 27.57 & 6.7  & 0 & 47.87     \\ \hline
\end{tabular}%
}
\label{tbl:sasa}
\end{table}
\smallskip
\noindent\textbf{Results of Multi-Floorplan Generation}

For HBM designs that are sensitive to logic resource utilization and routing resource utilization at the same time, we generate a set of Pareto-optimal floorplanning and implement all of them to explore the potentially best results. Table~\ref{tbl:incremental-opt} shows the corresponding achievable frequency. The number of generated floorplan candidates is related to the granularity of the design. Designs with larger tasks have less flexibility in floorplanning, thus there are fewer points on the Pareto-optimal curve. It remains as future work to automatically split large tasks and fuse small tasks to better facilitate the floorplan process.

As can be seen, even with the same set of optimization techniques, slightly different floorplanning may lead to non-trivial variation in the final achievable frequency. At this stage, we treat the downstream tools as a black box, so we implement all generated floorplan schemes in parallel to search for the best results. How to better predict the final frequency and skip unpromising floorplans in an early stage remains as future work.


\begin{table}[]
\caption{Comparison of different optimization levels. With the highest optimization level, we list the final frequency of all floorplan points.}
\resizebox{\columnwidth}{!}{%
\begin{tabular}{lllll}
\hline
Design  & Baseline & Multi-Floorplan Generation                                          & Max     & Min     \\ \hline
SASA    & 169 MHz  & 230 / 236 / 241 / Failed / Failed / 168 MHz                         & 241 MHz & Failed  \\
SpMM    & 216 MHz  & 242 / 232 / 231 / 238 / 194 / 215 / 245 MHz                         & 245 MHz & 194 MHz \\
SpMV-24 & 194 MHz  & 253 / 257 / 276 / 173 / 270 / 259 / 251 / 244 / 261 / 280 / 284 MHz & 284 MHz & 173 MHz \\
SpMV-16 & 233 MHz  & 300 / 285 MHz                                                       & 300 MHz & 285 MHz \\ \hline
\end{tabular}%
}
\label{tbl:incremental-opt}
\end{table}

\subsection{Control Experiments}

\begin{figure}[!h]
	\centering
	\includegraphics[width=0.8\columnwidth]{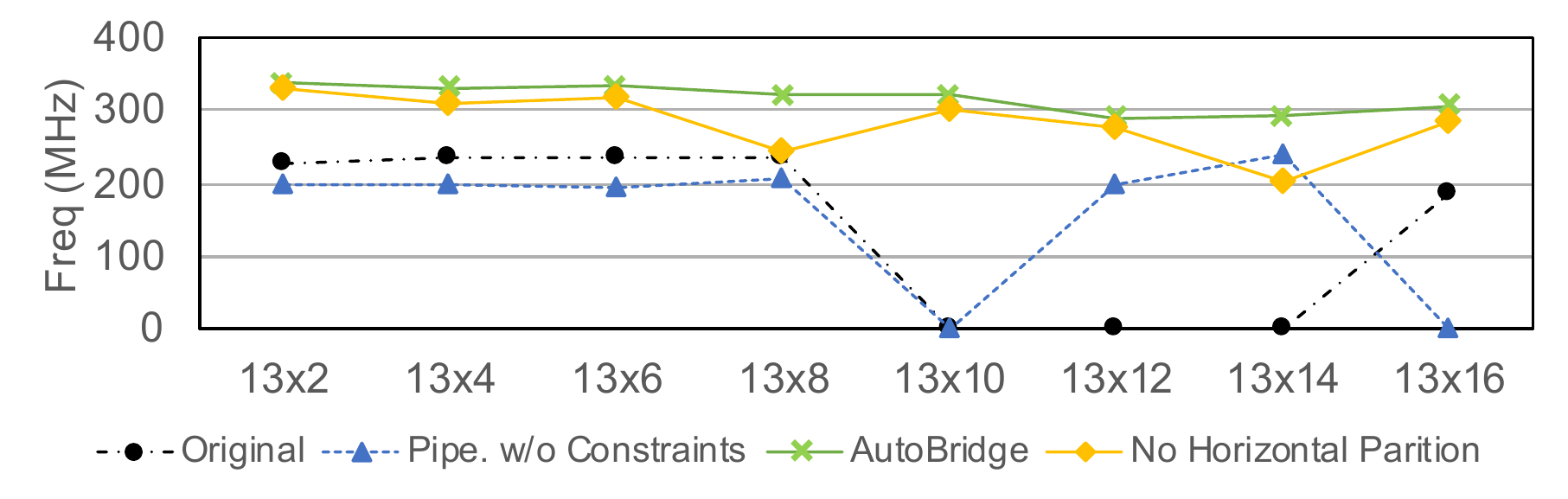}
	\caption{Control experiments with the CNN accelerators.}
    \label{fig:pipe_no_fp}
\end{figure}

First, we test whether the frequency gain comes from the combination of pipelining and HLS-floorplanning, or simply pipelining alone. To do this, we set a control group where we perform floorplanning and pipelining as usual, but we do not pass the floorplan constraints to the physical design tools. The blue curve with triangle markers in Figure~\ref{fig:pipe_no_fp} shows the results. As can be seen, the control group has a lower frequency than the original design for small sizes and has limited improvements over the original designs for large sizes. In all experiments, the group with both pipelining and floorplan constraints (green curve with crossing markers) has the highest frequency. This experiment proves that the frequency gain is not simply a result of more pipelining.

Meanwhile, if we only do floorplanning without pipelining, obviously the frequency will be much degraded, as visualized by Fig.~\ref{fig:polysa_dcp}.

Second, we test the effectiveness of setting a slot boundary based on the DDR controllers. We run a set of experiments where we only divide the FPGA into four slots based on the die boundaries, minus the division in the middle column. The yellow curve with diamond markers in Figure~\ref{fig:pipe_no_fp} shows the results. As can be seen, it achieves lower frequency compared to our default eight-slot scheme.  

\subsection{Scalability}

To show that the tool works well on designs with large numbers of small functions, we utilize the CNN experiments to test the scalability of our algorithms, as the CNN designs have the most vertices (HLS functions) and edges. Table~\ref{tbl:scalability} lists The compile time overhead for the floorplanning and the latency balancing when using Gurobi as the ILP solver\footnote{Meanwhile, we observed that many open-source ILP solvers are much slower.}. For the largest CNN accelerator that has 493 modules and 925 FIFO connections, the floorplan step only takes around 20 seconds and the latency balancing step takes 0.03s. Usually, FPGA designs are not likely to have this many modules and connections~\cite{url-vitis-library}~\cite{url-intel-opencl}, and our method is fast enough. 

\begin{table}[!h]
\caption{Computing time for the CNN test cases targeting the U250 FPGA. \textit{Div-1} and \textit{Div-2} denote the first and the second vertical decomposition, and \textit{Div-3} denotes the first horizontal decomposition. \textit{Re-balance} denotes the delay balancing. }
\resizebox{0.6\textwidth}{!}{%
\begin{tabular}{ccccccc}

\hline
Size  & \# V & \# E & Div-1 & Div-2 & Div-3 & Re-balance \\ \hline
13 $\times$ 2  & 87          & 141      & 0.02 s            & 0.02 s            & 0.01 s         & $<$0.01 s        \\
13 $\times$ 4  & 145         & 253      & 0.05 s            & 0.02 s            & 0.20 s            & $<$0.01 s        \\
13 $\times$ 6  & 203         & 365      & 0.07 s            & 1.02 s            & 0.56 s            & $<$0.01 s        \\
13 $\times$ 8  & 261         & 477      & 0.07 s            & 1.07 s            & 3.58 s            & 0.01 s           \\
13 $\times$ 10 & 319         & 589      & 3.17 s            & 1.61 s            & 2.63 s            & 0.01 s           \\
13 $\times$ 12 & 377         & 701      & 3.42 s            & 1.43 s            & 9.84 s            & 0.01 s           \\
13 $\times$ 14 & 435         & 813      & 3.54 s            & 1.55 s            & 6.18 s            & 0.03 s           \\
13 $\times$ 16 & 493         & 925      & 4.95 s            & 2.02 s            & 12.56 s           & 0.03 s          \\ \hline
\end{tabular}%
}
\label{tbl:scalability}

\end{table}

\section{Related Work}
\label{sec:related}



\noindent\textbf{Layout-Aware HLS Optimization.}
Previous works have studied how to couple the physical design process with HLS in a \textit{fine-grained} manner. Zheng~\textit{et~al.}~\cite{zheng2014fast} propose to iteratively run placement and routing for fine-grained calibration of the delay estimation of wires. The long run time of placement and routing prohibits their methods from benefiting large-scale designs, and their experiments are all based on small examples (1000s of registers and 10s of DSPs in their experiments).
Cong \textit{et. al.}~\cite{cong2004architecture} presented placement-driven scheduling and binding for multi-cycle communications in an island-style reconfigurable architecture. Xu \textit{et. al.}~\cite{xu1997layout} proposed to predict a register-level floorplan to facilitate the binding process. Some commercial HLS tools~\cite{url-cadence, url-synopsys} have utilized the results of logic synthesis to calibrate HLS delay estimation, but they do not consider the interconnect delays.

The previous approaches share the common aspect of focusing on the fine-grained interaction between physical design and upstream synthesis, where individual operators and the associated wires and registers are all involved during the delay prediction and iterative pipeline co-optimization. While such a fine-grained method can be effective on relatively small designs and FPGA devices, it is too expensive (if not infeasible) for today's large designs targeting multi-die FPGAs, where each implementation iteration may take days to complete.

In contrast, we focus on a coarse-grained approach that only pipelines the channels that span long distances and guides the detailed placement.

Other works have studied methods to predict delay estimation at the behavior level. Guo~\textit{et~al.}~\cite{guo2020analysis} proposed to calibrate the estimated delay for operators with large broadcast factors by pre-characterizing benchmarks with different broadcast factors. Tan~\textit{et~al.}~\cite{tan2015mapping} showed that the delay prediction of logic operations (e.g., \texttt{AND}, \texttt{OR}, \texttt{NOT}, etc) by HLS tools is too conservative. Therefore they consider the technology mapping for logic operations. These works mainly target local operators and have limited effects on global interconnects. Zhao~\textit{et~al.}~\cite{zhao2019machine} used machine learning to predict how manual pragmas  affect routing congestion. 

In addition, Cong~\textit{et~al.}~\cite{cong2018latte} presented tools to allow users to insert additional buffers to the designated datapath. Chen~\textit{et~al.}~\cite{chen2019high} proposed to add additional registers to the pipelined datapath during HLS synthesis based on the profiling results on the CHStone benchmark. \cite{zhang2019frequency} proposes to generate floorplanning constraints only for systolic array designs, and their method does not consider the interaction with peripheral IPs such as DDR controllers.
In comparison, our work is fully-automated for general designs and our register insertion is accurate due to HLS-floorplan co-design.

\smallskip
\noindent\textbf{Optimization for Multi-Die FPGAs.}
To adapt to multi-die FPGAs, previous works have studied how to partition the entire design or memories among different dies~\cite{roy1993timing, cherabuddi1994automated, mao2016modular, hahn2014cad, nasiri2015multiple, karypis1999multilevel, voss2019memory}. These methods are all based on RTL inputs, thus the partition method must observe the cycle-accurate specification. 
\cite{hahn2014cad, nasiri2015multiple} try to modify the cost function of placement to reduce die-crossing. This will lead to designs confined in fewer dies with a higher level of local congestion. Zha \textit{et al.}~\cite{zha2020virtualizing} propose methods to virtualize the FPGA and let different applications execute at different partitions. Xiao \textit{et al.}~\cite{xiao2020fast,xiao2022pld} propose methods to split the placement and routing of different parts of the design through dynamic reconfiguration.



\smallskip
\noindent\textbf{Floorplanning Algorithms.} 
Floorplanning has been extensively studied~\cite{alpert2008handbook,cheng2006floorplan,banerjee2009fast,murray2015hetris}. Conventionally, floorplanning consists of 1) feasible topology generation and 2) determining the aspect ratios for goals such as minimal total wire length. In the existing FPGA CAD flows, the floorplanning step works on RTL input. In contrast, we propose to perform coarse-grained floorplanning during the HLS step to help gain layout information for the HLS tool.
Similar to~\cite{lauther1988min,la1986mason,modarres1987automatic}, our algorithm adopts the idea of the partitioning-based approach. As our problem size is relatively small, we use ILP for each partitioning.

\smallskip
\noindent\textbf{Throughput Analysis of Dataflow Designs.}
Various dataflow models have been proposed in other literature, such as the Kahn Process Network (KPN)~\cite{gilles1974semantics}, Synchronous Data Flow (SDF)~\cite{lee1987synchronous}, among many others. The more simplified the model is, the more accurately we can analyze its throughput. In the SDF model, it is restricted that the number of data produced or consumed by a process for each firing is fixed and known. Therefore, it is possible to analytically compute the influence of additional latency on throughput~\cite{ghamarian2006throughput}. The latency insensitive theory (LIT)~\cite{carloni2000performance,lu2003performance,lu2006performance,collins2007topology,abbas2018latency} also enforces similar restrictions as SDF. \cite{venkataramani2014system} proposes methods to insert delays when composing IP blocks of different latency. \cite{josipovic2020buffer} studies the buffer placement problem in dataflow circuits~\cite{josipovic2018dynamically,cheng2020combining,cheng2022finding,cheng2021dass}. Other works have studied how to map dataflow programs to domain-specific coarse-grained reconfigurable architectures~\cite{weng2020dsagen,weng2020hybrid,dadu2019towards,dadu2021polygraph}.

In our scenario, each function will be compiled into an FSM that can be arbitrarily complex, thus it is difficult to quantitatively analyze the effect of the added latency on the total execution cycles. We adopt a conservative approach to balance the added latency on all reconvergent paths. 






\section{Future Work}

While TAPA has already improved significantly on the expressiveness and timing closure, there is a myriad of opportunities to further advance the tool. We list several challenges that we aim to address in the future.
\begin{itemize}
    \item Reduce the compile time by integrating RapidStream~\cite{guo2022rapidstream}. RapidStream builds on top of the idea of HLS-floorplanning co-optimization, and it further splits the design for parallel placement and routing. When tested on the Xilinx U250 FPGA with a set of realistic HLS designs, RapidStream achieves a 5-7$\times$ reduction in compile time and up to 1.3$\times$ increase in frequency when compared to a commercial-off-the-shelf toolchain. We are in the progress to integrate RapidStream with TAPA.
    \item Support more flexible inter-task communication patterns. Currently, TAPA tasks can only communicate with each other through streams. We are extending the infrastructure to support buffer-based channels between tasks for richer expressiveness. 
    \item Task-level compiler optimization. As of now, TAPA delegates the compilation of each task to existing HLS tools and does not perform inter-task optimizations. This limitation requires that users come up with a good partitioning of the application into tasks of suitable sizes. We aim to add additional task-level optimization such as task splitting, task fusion, task hierarchy rebuild, etc. to further co-optimize the task hierarchy and the floorplanning process. 
    \item Support designs with a hybrid of RTL and HLS tasks. The floorplan-guided pipeline methodology could apply to RTL tasks as long as they adhere to pre-defined latency-insensitive interfaces. Although we have explored applying the technique to RTL projects~\cite{qiao2021fans, qiao2022topsort}, more efforts are needed to provide an automated solution.
\end{itemize}

\section{Conclusions}

In this paper, we present the TAPA framework, an efficient task-parallel programming tool for modern FPGAs. TAPA includes a set of convenient APIs to increase design efficiency. In addition, we tackle the challenge of high-frequency HLS design on multi-die FPGAs by coupling floorplanning and pipelining to effectively insert registers on the long cross-die interconnects. We present a set of optimizations specifically tailored for HBM devices, including automatic HBM port binding, floorplan solution space exploration, and a customized programming API to minimize the area overhead of HBM IO modules. Our framework, TAPA, interfaces with the commercial FPGA design tool flow. It improves the average frequency of 43 designs from 147 MHz to 297 MHz with a negligible area overhead. TAPA has been used in multiple projects to improve the design efficiency and final frequency~\cite{chi2022accelerating, wang2021autosa, song2022serpens, song2022sextans, dac20-soda-cr}.



\begin{acks}
{\small
We greatly thank Wuxi Li, Jianyi Cheng, Luciano Lavagno, Gai Liu, Zixuan Jiang, Yifan Yuan and the anonymous reviewers for their valuable feedback.
This work is partially supported by the CRISP Program, members from the CDSC Industrial Partnership Program, the Intel/NSF CAPA program, the NSF NeuroNex Award No. DBI-1707408, the NIH Award No. U01MH117079, the NSERC Discovery Grant RGPIN-2019-04613 and DGECR-2019-00120, and the CFI John R. Evans Leaders Fund. The authors acknowledge the valuable support of the Xilinx Adaptive Compute Clusters (XACC) Program. We thank Gurobi and GNU Parallel for their support to academia.
}
\end{acks}

\bibliographystyle{ACM-Reference-Format}
\bibliography{main}

\end{document}